\documentclass[twocolumn,dvipsnames]{aastex631}
\usepackage[T1]{fontenc}
\usepackage{amsmath}
\usepackage{natbib}
\usepackage{soul}
\usepackage{xcolor}
\usepackage{lipsum}
\usepackage{graphicx}
\usepackage{float}
\usepackage{graphicx}	
\usepackage{multirow,makecell}
\usepackage{placeins}
\usepackage{bm,amsfonts,hhline,amssymb,microtype,float,latexsym,epsf,mathtools,times,makecell,array}
\usepackage{makecell}
\usepackage{tabularx}
\usepackage{epstopdf}
\usepackage{gensymb}
\usepackage{pifont}
\usepackage{hyperref}
\hypersetup{colorlinks=true, linkcolor=red, filecolor=magenta, urlcolor=Sepia, citecolor=blue}

\definecolor{ForestGreen}{HTML}{228B22}

\received{XXX}
\revised{YYY}
\accepted{ZZZ}

\shorttitle{Continuous gravitational waves from magnetized white dwarfs}
\shortauthors{Das, Mukhopadhyay \& Bulik}

\begin{document}

\title{Continuous gravitational waves from magnetized white dwarfs: Quantifying the detection plausibility by LISA}

\author[0009-0004-6340-9337]{Mayusree Das$^{\dagger}$}
\email{mayusreedas@iisc.ac.in$^{\dagger}$}
\affiliation{Joint Astronomy Programme, Department of Physics, Indian Institute of Science, Bangalore, 560012, India}

\author[0000-0002-3020-9513]{Banibrata Mukhopadhyay$^{\ddagger}$}
\email{bm@iisc.ac.in$^{\ddagger}$}
\affiliation{Department of Physics, Indian Institute of Science, Bangalore, 560012, India}
\affiliation{Joint Astronomy Programme, Department of Physics, Indian Institute of Science, Bangalore, 560012, India}

\author[0000-0003-2045-4803]{Tomasz Bulik$^{*}$}
\affiliation{Astronomical Observatory, University of Warsaw, Al. Ujazdowskie 4, 00478 Warszawa, Poland}
\email{tb@astrouw.edu.pl$^{*}$}

\begin{abstract}

  White dwarfs (WDs) are frequently observed to have strong magnetic fields up to $10^9$ G and expected to have a possible internal field as high as $\sim 10^{14}$ G. High internal fields can significantly deform a WD's equilibrium structure, generating a quadrupole moment. If the rotation axis is misaligned with the magnetic axis, the deformation can lead to the emission of continuous gravitational waves (CGWs). We examine the potential for detecting CGWs from magnetized WDs with future space-based detectors such as LISA, ALIA, DECIGO, Deci-Hz, BBO and TianQin. We model the field-induced deformation and compute the resulting GW strain, incorporating amplitude decay due to angular momentum loss from electromagnetic and gravitational radiation. This sets a timescale for detection -`active timescale' of $10^{5-6}$ yr, requiring observation while the object remains sufficiently young. Our results suggest that LISA could detect a few dozens of highly magnetized WDs across the Galaxy during its mission. As a specific case, we investigate ZTF J1901+1458- a compact, massive, fast-rotating, and strongly magnetized WD with spin period $\sim416$ s and inferred surface field $\sim10^{9}$ G. We find that this object would be detectable by LISA with four years of continuous data. This highlights the potential of CGW observations to probe magnetic field structure in WDs and their role in type Ia supernova progenitors.

\end{abstract}

\section{Introduction}

A substantial number of white dwarfs (WDs) display clear signatures of magnetic fields. Observations using techniques such as Zeeman spectroscopy and polarimetry---particularly from large-scale surveys like the Sloan Digital Sky Survey (SDSS) \citep{Alam2015}---indicate that around 5--20\% of WDs possess high surface magnetic field strengths ranging from $10^6$ to $10^9$~G \citep{Ferrario2015, Ferrario2020}. 

The strong magnetic fields observed in WDs may originate from their main-sequence progenitors and be enhanced through stellar evolution via magnetic flux freezing \citep{wick2005}. Alternatively, these fields could be generated toward the end of the main-sequence phase and maintained during the WD formation. In binary systems, particularly double WD systems, magnetic fields might also be produced by dynamo processes driven by binary interactions \citep{tout2008}. It is also possible that the seed magnetic field in the WD's initial stage is enhanced by turbulence driven dynamo during WD's crystallization \citep{Isern2017}.

 Strong magnetic fields significantly affect the hydrostatic equilibrium of WDs, leading to deviations from spherical symmetry. The resulting deformation depends on both the strength and configuration of the internal magnetic field, and is generally proportional to the magnetic energy content of the star \citep{Chandrasekhar1953}. Consequently, WDs with stronger internal magnetic field $\sim 5\times 10^{13}$ G are expected to exhibit greater distortion. Mukhopadhyay and collaborators have shown that when such a deformed WD rotates about an axis misaligned with its magnetic axis, it can emit gravitational waves (GWs), analogous to what is proposed for neutron stars (NSs) \citep{KM2019,sold2021,das-mukhopadhyay,gian2025}. This type of emission arises from a time-varying quadrupole moment. Even though the system is isolated, it produces a long-lived signal integrated over the observing time as long as the object remains magnetized and continues to rotate with non-zero misalignment between magnetic and rotation axes, calling obliquity angle $\chi$. This GW signal is therefore referred to as a continuous gravitational wave (CGW) in the standard gravitational-wave nomenclature (see \citealt{Riles2023} for a review), emitted from isolated WDs in mHz band.

\begin{table*}
\begin{center}
\caption{Magnetized WDs with $\chi=3^\circ$ and $d=1$ kpc. The surface field is orders of magnitude smaller than $B_{max}$.}
\small
\begin{tabular}{c c c c c c c c c c c c} 
\hline\hline
Configuration & $\rho_c$ (g cm\textsuperscript{-3}) & $M$ $(M_{\odot})$&$R_E$ (km) & $B_{max}$ (G) & $\nu$ (Hz) & ME/GE & $|\epsilon|$ & $h(t=0)$ \\
\hline
WD1:Toroidal & $10^9$ & 1.83 & 5708 & $3.9\times 10^{13}$ & 0.05 & 0.15 & 0.54 & $1\times 10^{-23}$ \\
WD2:Poloidal & $10^9$ & 1.42 & 2815 & $4\times 10^{13}$ & 0.05 & 0.05 & 0.17 & $5\times 10^{-25}$ \\
WD3:Toroidal & $10^9$ & 1.4 & 3534 & $2.5\times 10^{13}$ & 0.05 & 0.04 & 0.11 & $6\times 10^{-25}$ \\
\hline
\end{tabular}
\label{tab:tableblind}
\end{center}
\end{table*}

The prospect of detecting magnetized WDs through CGWs becomes highly promising with the advent of future detectors like LISA, TianQin. However, the amplitude of these GWs diminishes over time as the system loses angular momentum through both electromagnetic and gravitational radiation. This natural spin-down imposes a timescale constraint—magnetized WDs must be observed while they are still young and the signal remains above detector sensitivity thresholds. In this work, we explore the feasibility and probability of detecting such sources in all sky searches with future space-based GW observatories including LISA \citep{LISAdefinitionstudy}, DECIGO \citep{Kawamura2020}, Deci-Hz (Rajesh K. Nayak et al., private communication) \footnote{Proposed Indian GW detector in Deci-Hz frequency; based on its representative noise PSD.}, ALIA \citep{Crowder2005}, BBO \citep{CutlerHarms2006} and TianQin \citep{TianQinoverview}. Intriguingly, we highlight that the recently discovered ZTF J1901+1458- one of the most compact, near-Chandrasekhar mass, rapidly rotating and strongly magnetized WDs \citep{ZTF} known—stands out as a particularly exciting and potentially detectable CGW emitter for LISA targeted search.

Highly magnetized, fast rotating WDs have been {theoretically shown} to potentially exceed the standard Chandrasekhar mass limit \citep{dasmuk13, Das2014}. {If such WDs happen to be accreting from a nondegenerate companion}, and approach their (super-Chandrasekhar) limiting mass (e.g. \citealt{dasmuk13,deb2022}), they can no longer be supported by electron degeneracy pressure{. This} may eventually trigger an overluminous type Ia supernova. In particular, detecting these massive {(even significantly) more than $1.4 M_\odot$}, magnetized WDs through CGW could help in understanding the origins of overluminous type Ia supernovae, such as SN 2003fg \citep{howell2006}.

{In this work, we connect the theoretical modeling of magnetically deformed rotating WDs with their direct detectability through CGWs, quantifying their detection plausibility with upcoming detectors and including a study for targeted detection.} The structure of this paper is as follows. In Section \ref{White dwarf model}, we present the numerically simulated models of deformed, rotating, and magnetized WDs with a nonzero $\chi$, which serve as potential sources of GW emission. Section \ref{GW decay} outlines the physical processes responsible for the gradual reduction in GW amplitude over time setting a timescale for detection while they are young. We assess the likelihood of detecting such signals in Section \ref{GWdetect}, where we calculate the Signal-to-Noise ratio in Section~\ref{snr}. In Section \ref{Detection plausibility} we estimate the expected number of WDs to be detected by LISA once operating, taking into account the decay timescale and birthrate of WDs. The detectability of CGWs from the specific case of ZTF J1901+1458 using LISA is discussed in Section \ref{ztf}. In Section \ref{gwdist}, we discuss how to distinguish these CGWs from the Galactic binary background. Finally, we provide a conclusion of our findings in Section \ref{Conclusion}.

\section{White dwarf model}
\label{White dwarf model}

\begin{figure}
\begin{center}
\includegraphics[scale=0.6]{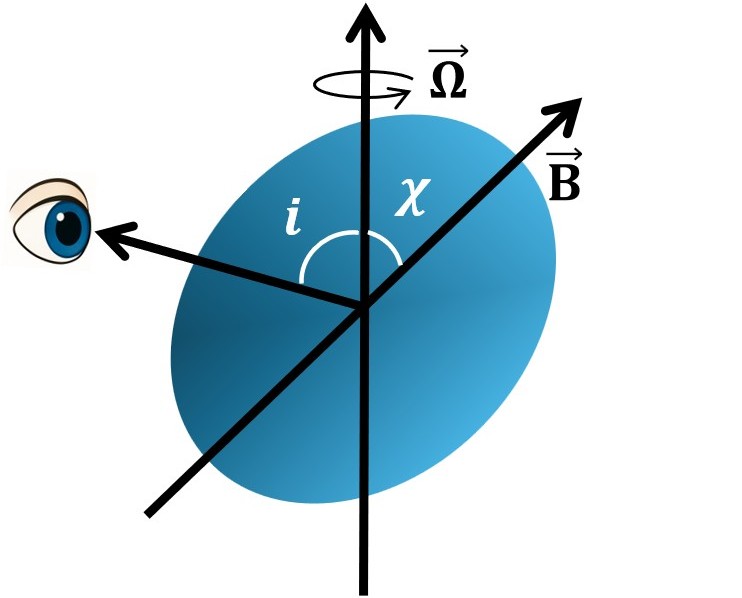}
\caption{A cartoon diagram of magnetized rotating WD
with misalignment between the magnetic field axis with and the rotation axis. Here $i$ is the angle between the rotation axis of the source and detector, $\chi$ is the misalignment between magnetic and rotation axes}
\label{fig:cartoon}
\end{center}
\end{figure}

\begin{figure*}
\begin{center}
\includegraphics[scale=0.9]{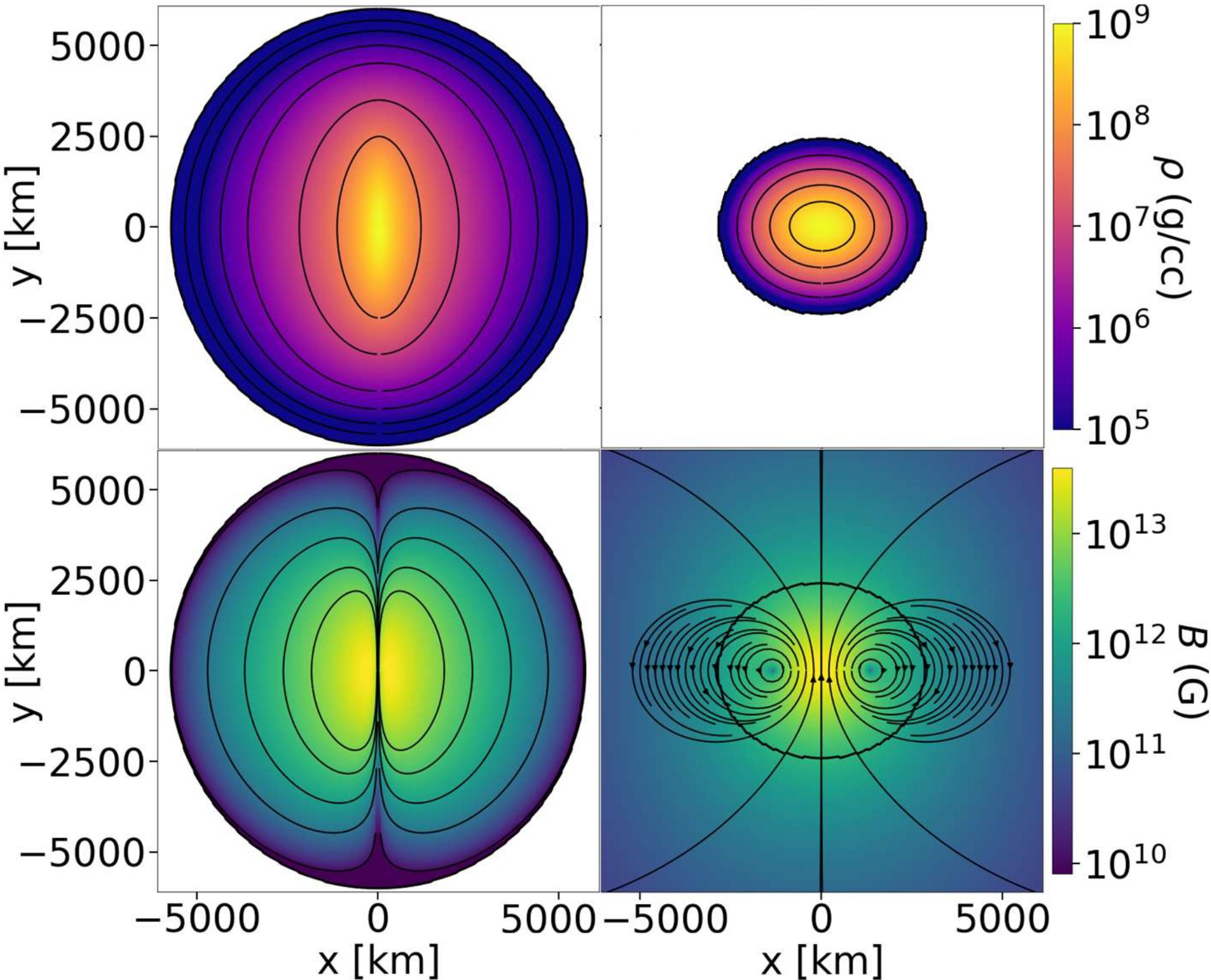}
\caption{Density (top panels) and magnetic field (bottom panels) are shown in colorbar for WD1 (of $M=1.83\,M_\odot$): toroidal (left panels) and WD2 (of $M=1.42\,M_\odot$): poloidal (right panels). The isocontour lines are drawn for density, toroidal magnetic field (having magnetic field perpendicular to the plane) and the poloidal magnetic field lines with arrow. The detailed properties of those WDs are mentioned in Table. \ref{tab:tableblind}}.
\label{fig:xns}
\end{center}
\end{figure*}

In the context of WDs, a non-zero time-varying quadrupole moment can arise from triaxial deformations induced by strong internal magnetic fields. The poloidal and toroidal field components exert anisotropic stresses that deform the stellar structure into oblate or prolate configurations, respectively. However, these deformations are axisymmetric with respect to the magnetic symmetry axis and GW emission arises only when $\chi \neq 0$. Indeed a nonzero $\chi$ is generally expected in magnetized WDs formed by merger or accretion \citep{GarciaBerro2012,Ji2013,Ferrario2015}. The degree of deformation is quantified by the ellipticity, defined as $\epsilon = |I_{zz} - I_{xx}| / I_{xx}$ for axisymmetric system, where $I_{ij}$ are the components of the moment of inertia tensor and $\hat{z}$ is aligned with the rotation axis of the WD. When the rotation axis is not aligned with the magnetic symmetry axis, the resulting configuration leads to a time-varying quadrupole moment. This misalignment $\chi$ as shown in Figure~\ref{fig:cartoon}- a schematic illustration of rotating, magnetized WD. is essential for CGW emission, as it introduces periodic variations in the mass distribution from the observer's frame. If the WD rotates with angular frequency $\Omega=2\pi\nu$, the GW strain will have contributions at frequencies $\Omega$ and and its harmonic $2\Omega$, arising due to quadrupole nature. The
 corresponding two polarization modes of the dimensionless gravitational wave strain are given by \citep{BG1996}
\begin{equation}
\begin{aligned}
h_+ &= h_0 \sin \chi \left[ 
       \tfrac{1}{2} \cos i \sin i \cos \chi \cos \Omega t \right. \\
    &\qquad \qquad\qquad\qquad\qquad\quad\left. - \tfrac{1+\cos^2 i}{2} \sin \chi \cos 2 \Omega t \right], \\
h_\times &= h_0 \sin \chi \left[ 
       \tfrac{1}{2} \sin i \cos \chi \sin \Omega t 
       - \cos i \sin \chi \sin 2 \Omega t \right],
\end{aligned}
\label{eq:gwstrain}
\end{equation}
where  $i$ is the angle between the rotation axis of the source and detector. Here the GW amplitude is
\begin{equation}
    h_0=\frac{2G}{c^4} \frac{\Omega^2 \epsilon I_{xx}}{d}\left(2\cos^2\chi-\sin^2\chi\right),
\label{eq:gwh0}
\end{equation}
where $c$ is the speed of light, $G$ is Newton's gravitational constant, $d$ is the distance between the detector and the source. The GW strain averaged over the time period $T=2\pi/\Omega$ of the WD is
\begin{equation}
\left\langle h \right\rangle = \sqrt{ \left\langle h_+^2 \right\rangle_T + \left\langle h_\times^2 \right\rangle_T},
\label{eq:h_avg_root}
\end{equation}
where
\begin{align}
\left\langle h_+^2 \right\rangle_T &= \frac{h_0^2 \sin^2\chi}{8} \left[ \cos^2 i \sin^2 i \cos^2 \chi \right. \notag \\[1ex]
&\left. \quad + (1 + \cos^2 i)^2 \sin^2 \chi \right], \label{eq:hplus_avg} \\
\left\langle h_\times^2 \right\rangle_T &= \frac{h_0^2 \sin^2\chi}{2} \left( \frac{1}{4} \sin^2 i \cos^2 \chi + \cos^2 i \sin^2 \chi \right). \label{eq:hcross_avg}
\end{align}
Replacing the $h_0$ from Equation \eqref{eq:gwh0} into Equation \eqref{eq:h_avg_root} we obtain,
\begin{align}
\left\langle h \right\rangle &=  \frac{2G}{c^4} \frac{\Omega^2 \epsilon I_{xx}}{d}
\left(2\cos^2\chi-\sin^2\chi\right) \frac{\sin\chi}{2\sqrt{2}} \notag \\
&\quad\left[ 
\cos^2 i \sin^2 i \cos^2 \chi + \sin^2 i \cos^2 \chi \right. \notag \\
&\quad \left. + (1 + \cos^2 i)^2 \sin^2 \chi + 4\cos^2 i \sin^2 \chi 
\right]^{1/2}.
\label{eq:h_avg_formula}
\end{align}

To assess the feasibility of detecting WDs through CGW, it is essential to estimate the strain amplitude $h$, which depends on the degree of quadrupolar deformation caused by magnetic fields. This, in turn, requires knowledge of the moment of inertia along different principal axes, i.e. $I_{zz}$ about the rotation axis $\hat{z}$ as illustrated in Figure~\ref{fig:cartoon} and $I_{xx}$ about the perpendicular axis. Obtaining such detailed internal information observationally would require extremely high-resolution optical data, which is not yet available.

Alternatively, one can model these deformations using numerical simulations. For this purpose, we employ the general relativistic, steady-state, magneto-hydrostatic solver XNS 4.0 \citep{sold2021main}, originally developed for NS configurations. We have adapted and modified this code to simulate rotating, magnetized WDs. The interior of WD is described using a relativistic equation of state for a highly degenerate electron gas, approximated by the polytropic relation between pressure $P$ and density $\rho$ as, $P = K\rho^{4/3}$, with $K = 4.8 \times 10^{14}$ in CGS units. Figure~\ref{fig:xns} presents 2D cross-sectional views and magnetic field configurations illustrating the resulted deformations. Such WDs are modeled with cental density $\rho_c$ and maximum magnetic field $B_{max}$ as inputs to the model for a WD with particular $\nu$. The simulated quantities such as mass $M$, equatorial radius $R_E$, magnetic to gravitational energy ratio (ME/GE), magnetic field generated $\epsilon$ and hence $h$ are mentioned in Table \ref{tab:tableblind}. The models are called WD1: with toroidal magnetic field and WD2: with poloidal magnetic field, both having frequency $\nu=0.05$ Hz. It can be visualized that the WD with toroidal field is prolate in shape, whereas the WD with poloidal field is oblate in shape. As shown in Table \ref{tab:tableblind}, WD1 turns out to be significantly super-Chandrasekhar. {On the other hand, WD2 is close to the conventional $1.4 \,M_\odot$ Chandrasekhar limit.}

{The possible} existence of super-Chandrasekhar WD, like WD1, has been demonstrated in the presence of high magnetic fields in a series of papers (e.g. \citealt{dasmuk13, Das2014}). In particular, the toroidal component of the magnetic field contributes an additional outward radial pressure (through the anisotropic magnetic stress tensor) that is not attainable via rotation or finite temperature effects alone. While uniform rotation has been shown to increase the maximum stable WD mass only up to $\sim 1.5\,M_\odot$ \citep{Boshkayev2013} (though differential rotation may lead the mass to $1.8\,M_\odot$; \citealt{Subramanian2015}), strong toroidal fields can push the equilibrium mass {beyond $1.8\,M_\odot$}. We have discussed the detection of WD1-like WDs to optimize GW detection (as $h \propto I$), owing to their large equatorial radius $R_E$. Later we have also discussed detection of sub-Chandrasekhar WD in targeted search.

In presence of magnetic deformation, the other key ingredient for time-varying quadrupole moment is the misalignment. However, XNS does not have a provision to include $\chi$; we assume its effect to be small. We extract $\epsilon$ from XNS to calculate $h$ with small $\chi$ approximation, so that the angle does not affect the geometry e.g. the equatorial radius of the star as it is modeled in XNS for zero misalignment. The equatorial radius is $\cos\chi$ times circular radius (see Figure \ref{fig:cartoon}), the latter can be assumed to be similar to the former as long as $\chi\lesssim  30\degree$ (i.e., $\cos\chi>0.9$).

Now we can calculate GW strain once we decide $d$, $\chi$ and $i$. We will assume $\chi=3\degree$ and $d=1$ kpc for blind all sky search. However, $i$ is not known for blind search and also for most targeted search for WDs. Hence, for our WD model, the GW strain should be averaged over all possible sky positions as 
\begin{align}
\left\langle h(\chi, \nu) \right\rangle_i 
&= \frac{1}{2} \int_{-1}^1 \left\langle h(i) \right\rangle \,d(\cos i) \notag \\
&= \frac{1}{2} \int_0^\pi \left\langle h(i, \chi, \nu) \right\rangle_T\, \sin i \, di \notag \\
&= f(\chi)\frac{2G}{c^4} \frac{4\pi^2\nu^2 \epsilon I_{xx}}{d},
\label{eq:h_chi_avg}
\end{align}
where $f(\chi)$ is a function of only $\chi$ after averaging over $i$.

\section{GW decay}
\label{GW decay}

\subsection{{Magnetic field decay}}
\label{bdecay}

\begin{figure}
\begin{center}
\includegraphics[width=\columnwidth]{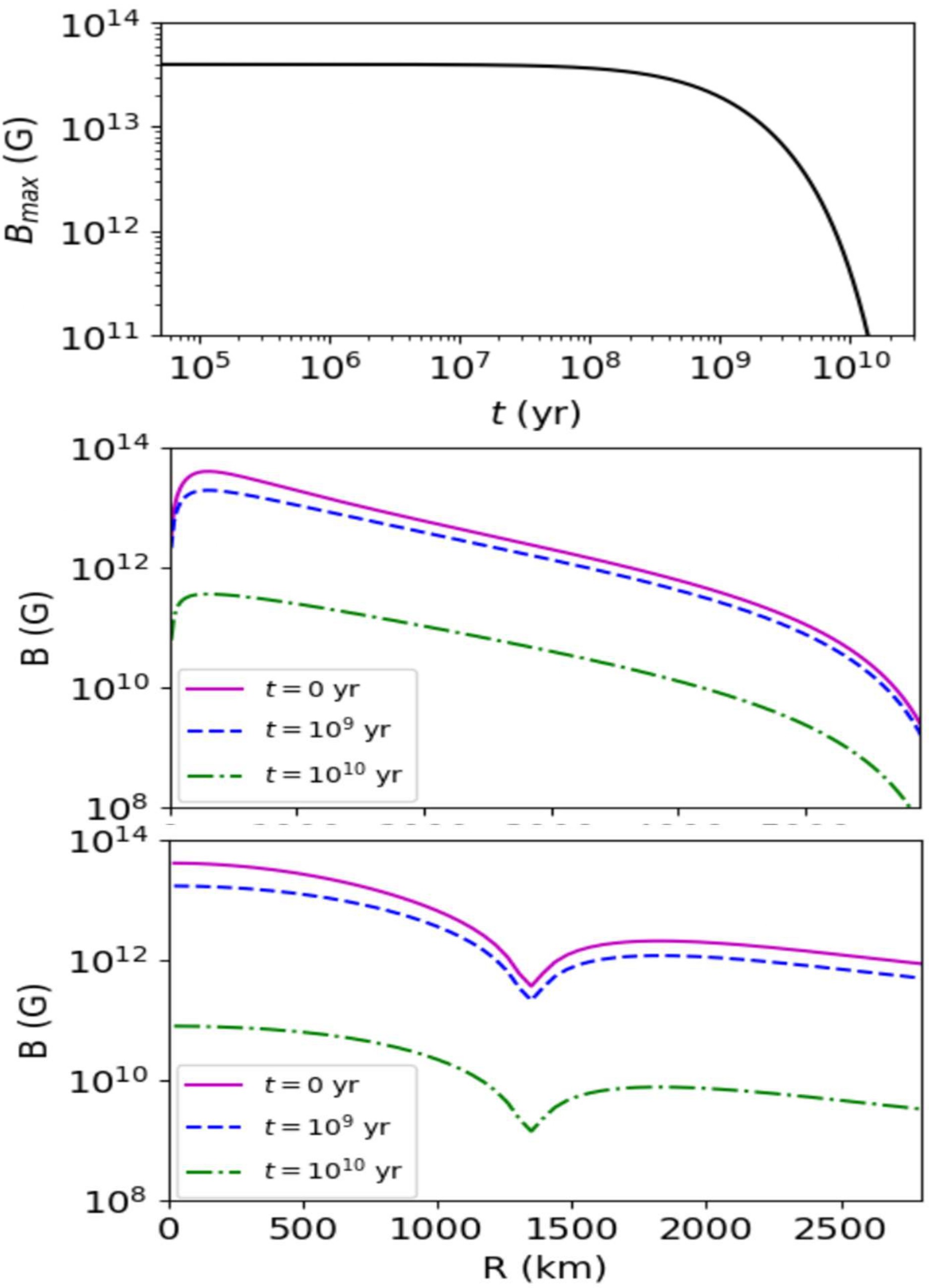}
\caption{Top panel: The decay of maximum magnetic $B_{max}=4\times 10^{13}$ G with time for WD1 and WD2 with radius $5708$ and $2815$ km, respectively. Middle: Magnetic field as a function of radius, before and after magnetic field decay for toroidal field (WD1). Bottom panel: Same as middle panel except for poloidal field (WD2)}.
\label{fig:bdecay}
\end{center}
\end{figure}

Strong internal magnetic field on the order of $10^{13-14}$ G, the key factor to deform the star and thus to produce CGW from WDs, decays with time due to Ohmic decay and Hall drift \citep{GR1992, HK1998}. The decay is given by
\begin{equation}
\frac{dB}{dt}=-B \left( \frac{1}{t_{ohmic}}+\frac{1}{t_{Hall}} \right),
\label{eq:Bdecay}
\end{equation}
where the timescales for Ohmic, and Hall decay can be found in \citealt{GR1992, HK1998,bhattacharya22}.
The initial magnetic field profile is executed from WD model. We show the evolution of magnetic fields for WD models (Table \ref{tab:tableblind}) in Figure \ref{fig:bdecay} where the timescale of decay is more than 10 Gyr for WD interiors. The evolutions of $B_{max}$ (top panel) for WD1 and WD2 coincide because both models were constructed with the same $\rho_c$ and $B_{max}=4\times10^{13}$ G. In this regime ($10^{12}$–$10^{14}$ G) the Hall timescale, which depends mainly on those two quantities, is much shorter than the Ohmic timescale and therefore governs the evolution, leading to identical $B_{max}$ decay. For the calculation of those timescales we assume same fiducial temperature $\sim 10^6$ K for all WDs.

\subsection{Angular momentum decay}
\label{spindown}

\subsubsection{Dipolar and GW radiation}
\label{dipolardecay}


A WD possessing a time-dependent, non-zero quadrupole moment will emit both electromagnetic dipole radiation and gravitational radiation due to quadrupolar energy loss. This results in angular momentum extraction from the system resulting in reduction of both $\Omega$ and  $\chi$, effectively spinning down the star and leading $\chi\rightarrow 0$. The corresponding dipole and quadrupole luminosities, are analogous to those used in the context of NSs in vacuum \citep{CH1970,KMMB2020,das-mukhopadhyay}. However, magnetized NSs as well as WDs are supposed to have an extended magnetosphere. In the presence of a magnetosphere, plasma-filled currents and open magnetic field lines enable more efficient extraction of rotational energy via Poynting flux, even for aligned rotators ($\chi\rightarrow0$). This enhanced energy outflow increases the spin-down luminosity compared to the vacuum case which was studied by \cite{Spitkovsky2006} for NS pulsars where the author calculated the dipole luminosity in presence of force free magnetosphere. The same description for luminosity can be used for WDs \footnote{WD's magnetosphere can be considered to be force free if plasma density $\rho_o<<{B^2}/{8\pi c^2}$. For WDs with surface dipolar field $10^9$ G, $\rho_o$ should be $<< 4\times 10^{-5}$ g cm\textsuperscript{-3}, which is indeed very common in isolated WD's magnetosphere. Even accreting system with accretion rate $\dot{M}=10^{-7}\, M_\odot/$yr, with light cylinder radius $r_{LC}=c/2\pi \nu\sim 3\times 10^{11}$ cm for $\nu=0.05$ Hz and free fall velocity at light cylinder $v_{ff}=\sqrt{2GM/r_{LC}}\sim 2\times 10^{7}$ cm s\textsuperscript{-1} can be considered as force free as $\rho_o=\dot{M}/(4\pi r_{LC}^2v_{ff})\sim 3\times 10^{-13}$ g cm\textsuperscript{-3} which is $<<4\times 10^{-5}$ g cm\textsuperscript{-3}.} by calculating magnetic moment ($=B_PR_P^3/2$) appropriately.

The electromagnetic (dipole) and GW (mass quadrupole) luminosities \citep{CH1970,Spitkovsky2006,KMMB2020,das-mukhopadhyay} are given by, respectively,
\begin{equation}
L_D\sim\frac{B_p^2 R_p^6 \Omega^4}{4c^3} (1+1.1\sin^2 \chi),
\label{eq:LD}
\end{equation}
and
\begin{equation}
L_{GW}=\frac{2G}{5c^5} (I_{zz}-I_{xx})^2 \Omega^6 \sin^2 \chi (1+15 \sin^2 \chi).
\label{eq:LGW}
\end{equation}
The corresponding evolution equations for $\Omega$ and $\chi$ are
\begin{align}
    \frac{d(\Omega I_{z^{\prime}z^{\prime}})}{dt}=&-\frac{2G}{5c^5} (I_{zz}-I_{xx})^2 \Omega^5 \sin^2 \chi (1+15 \sin^2 \chi) \nonumber \\
&-\frac{B_p^2 R_p^6 \Omega^3}{4c^3} (1+1.1\sin^2 \chi)
\label{eq:dwdt}
\end{align}
and
\begin{align}
   I_{z^{\prime}z^{\prime}}\frac{d\chi}{dt}=&-\frac{12G}{5c^5} (I_{zz}-I_{xx})^2 \Omega^4 \sin^3 \chi \cos \chi \nonumber \\
&-\frac{B_p^2 R_p^6 \Omega^2}{4c^3} \sin \chi \cos \chi,
\label{eq:dxdt}
\end{align}
where $B_p$ is the dipolar magnetic field strength of poloidal magnetic field, $R_p$ is the polar radius, $R_0$ is the mean radius of the WD. The first and second terms of Equations (\ref{eq:dwdt}) and (\ref{eq:dxdt}) are corresponding to GW and electromagnetic radiations, respectively. Equations (\ref{eq:dwdt}) and (\ref{eq:dxdt}) are solved simultaneously to determine the time evolutions of $\nu$ and $\chi$. To solve them, we input the initial values of $I_{xx}$, $I_{zz}$, $B_p$, and $R_p$, which are extracted from WD models.

\begin{figure}
\begin{center}
\includegraphics[width=\columnwidth]{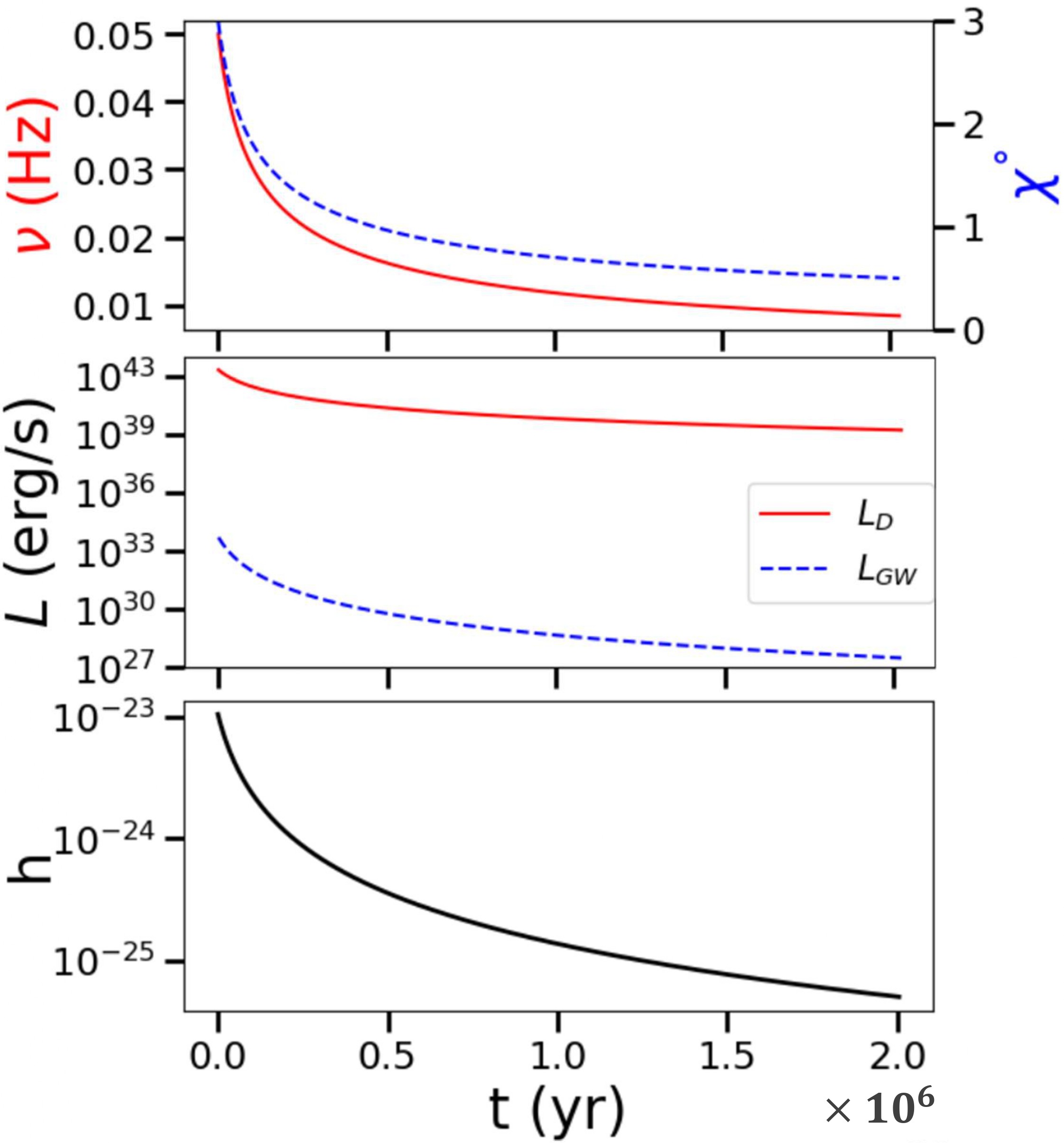}
\caption{Top Panel: Decay of $\nu$ (red solid), $\chi$ (blue dashed) as functions of time for initial $\nu=0.05$ Hz, $\chi=3^\circ$ for toroidally dominated WD- WD1+$B_P$ ( $B_{max}^{Toroidal}=4\times10^{13}$ G, $B_{p}^{Poloidal}=8\times 10^{8}$ G) with magnetosphere. For the same WD, Middle panel shows variation of luminosities and bottom panel shows GW strain decay with time.}
\label{fig:wdtorpolmagsp}
\end{center}
\end{figure}

\citet{KMMB2020} showed that $\nu$ and $\chi$ of WDs with purely toroidal magnetic field (similar to WD1 here) can decay significantly in $10^8$ yr, where gravitational radiation—driven by the quadrupole deformation—dominates the spin-down ($L_{GW}>>L_D$) as $B_p=0$. On the other hand, for purely poloidal fields (similar to WD2 here), electromagnetic losses significantly accelerate the evolution with $L_D>>L_{GW}$, leading to a rapid decay of $\nu$ and $\chi$ within just $10$ yr. However, it is well known that purely toroidal or poloidal stellar fields are inherently unstable. Consequently, mixed field models with toroidally dominated magnetic fields have been identified as stable in NSs and main sequence stars \citep{BR2009}. As WDs are the end states of some main sequence stars, it is expected that they follow a similar trend. Since the XNS code cannot model toroidally dominated mixed configurations, we assume a realistic WD, naming WD1+$B_P$, based on WD1 (which has a purely toroidal field; see Table~\ref{tab:tableblind}), but with a subdominant poloidal field of $B_p = 8 \times 10^8$~G -- an order of magnitude smaller than its surface toroidal field ($8 \times 10^9$~G; see Figure~\ref{fig:xns}) -- which does not affect the WD's structure. The choice of a surface poloidal field strength $\sim 10^{8-9}\,{\rm G}$ is motivated by recent observations from Zeeman spectroscopy and polarimetry \citep{Ferrario2015, Ferrario2020}. Such a poloidal component is expected to thread the magnetosphere and extend into the stellar interior, coexisting with an order-of-magnitude stronger internal toroidal field in order to ensure a stable magnetic configuration.

\begin{figure}
\begin{center}
\includegraphics[width=\columnwidth]{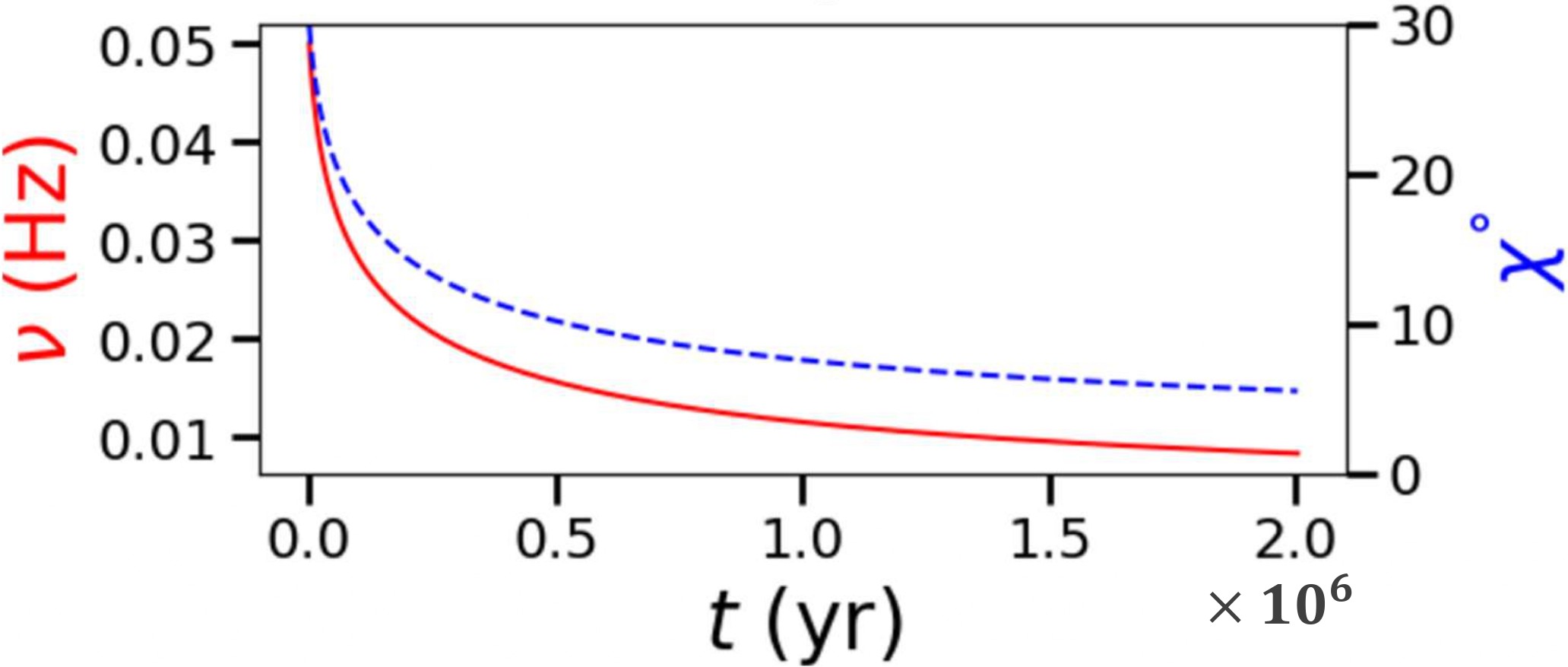}
\caption{Same as top panel of Figure \ref{fig:wdtorpolmagsp}, except with initial $\chi=30\degree$.}
\label{fig:torwxlgw30}
\end{center}
\end{figure}

\begin{figure}
\begin{center}
\includegraphics[width=\columnwidth]{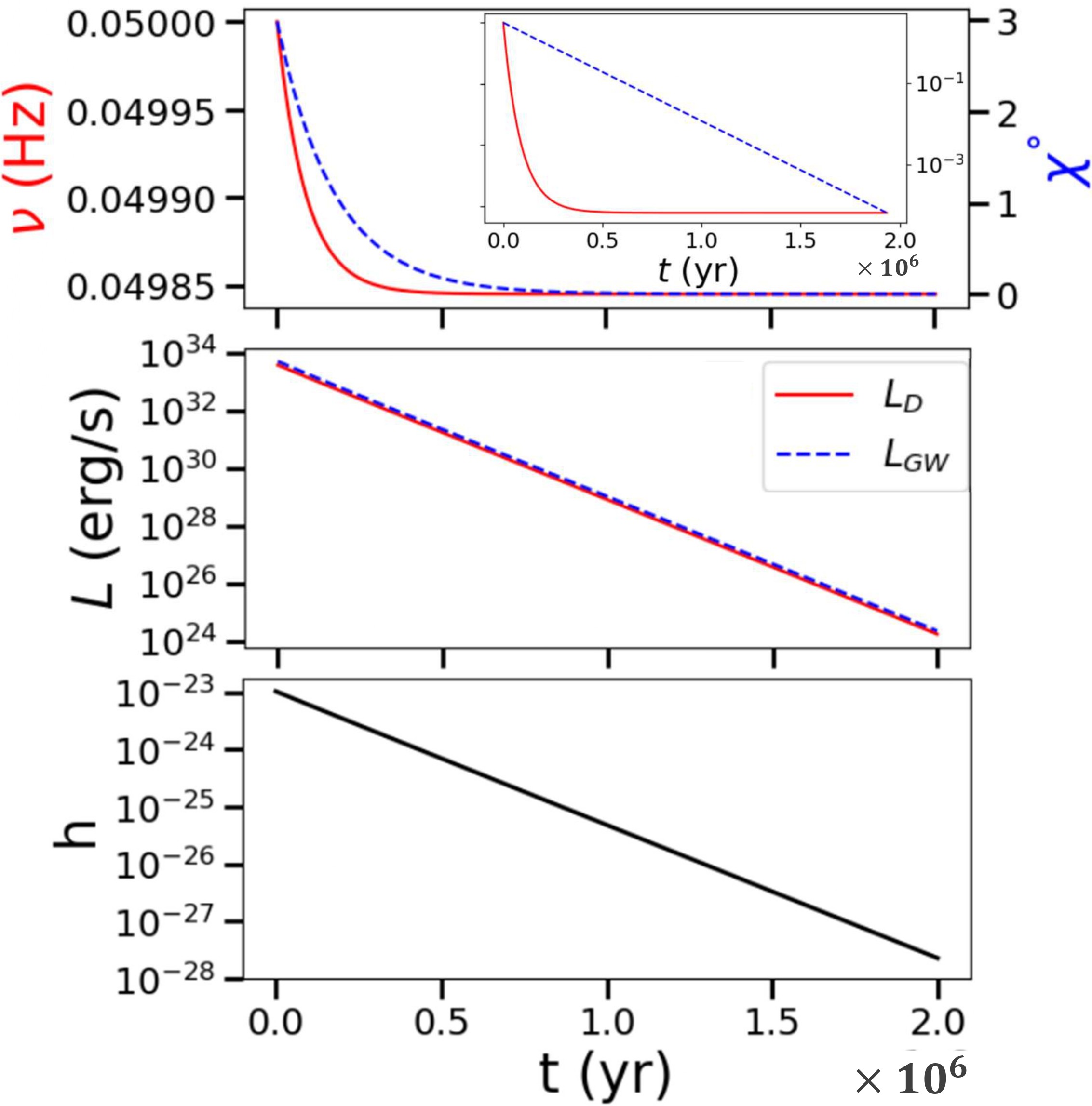}
\caption{Same as Figure \ref{fig:wdtorpolmagsp} for  WD1+$B_P$, except without magnetosphere. As $\chi\rightarrow 0$, the decays are shown in log scale too in inset plot of top panel.}
\label{fig:wdtorpol}
\end{center}
\end{figure}

For WD1+$B_P$, with magnetosphere, we show $\nu$ and $\chi$ decay, and luminosities in Figure \ref{fig:wdtorpolmagsp} (top and middle panels, respectively). The decay timescale reduces to $\sim 10^6$~ yr—shorter than that of a purely toroidal WD (where $L_D>>L_{GW}$) because of additional dipole effect. Here $\chi$ decays but saturates at $\chi\ne 0$ as $\nu$ decays because of enhanced energy outflow by magnetosphere.

 We show the decays of $\nu$ and $\chi$ in Figure \ref{fig:torwxlgw30} for another WD, with parameters same as WD1+$B_P$ except with initial $\chi=30\degree$. The timescale of decay and saturation values of $\nu$ and $\chi$ remain almost similar to Figure \ref{fig:wdtorpolmagsp} (top panel), concluding that the initial $\chi$ practically does not effect the decay timescale.

For comparison with the case including a magnetosphere (Figure~\ref{fig:wdtorpolmagsp}), we show in Figure~\ref{fig:wdtorpol} the evolution for WD1+$B_P$ without a magnetosphere. In this case the spin frequency $\nu$ hardly decays, while $\chi$ continues to decrease, leading to faster alignment. When $\chi \rightarrow 0$, the alignment occurs on the timescale $\sim 5\times 10^5$ yr, significantly shorter than in the magnetosphere case.

We notice that the GW strain $h$ is not constant; rather, it is a function of $\epsilon$, $\nu$, $\chi$ (Equation \ref{eq:h_chi_avg}), which decays with time; $\epsilon\propto$ magnetic field which has a decay timescale of $10^{10}$ yr, hence, we consider $\epsilon$ be constant throughout the evolution. However, due to angular momentum extraction from the system, $\nu$, $\chi$ and thus $\left\langle h(\chi, \nu) \right\rangle_i$ decay. This is shown in bottom panels of Figures \ref{fig:wdtorpolmagsp} and \ref{fig:wdtorpol}, for WDs with and without magnetosphere, respectively. In the magnetosphere case, the spin frequency $\nu$ decays strongly by several factors over $\sim 10^{6}$ yr, and $h$ is reduced by nearly two orders of magnitude. In the vacuum case without a magnetosphere, $\nu$ changes very little, but $h$ declines rapidly. Thus, WDs with magnetospheres evolve as sustainable long-lived CGW emitters, whereas those without experience a faster suppression of GW strain despite retaining high $\nu$.

\subsubsection{Multipolar and GW radiation}

Multipolar magnetic fields are increasingly supported in WDs by both observations and theory. Spectropolarimetry and Zeeman tomography reveal surface field structures often inconsistent with pure dipoles \citep{Euchner2002, euchner2005,Bagnulo2023}, suggesting presence of higher order multipolar field. If the internal magnetic field of the star contains higher-order multipolar components, it can be modeled as a point multipole characterized by its order $\ell$ (such that $\ell = 1$ for dipole, $\ell = 2$ for quadrupole, etc.) and azimuthal mode $m$ ($0\rightarrow l$), which together define the geometry of the field \citep{Petri2015}. These indices correspond to the decomposition of the electromagnetic field in terms of vector spherical harmonics. \citet{Petri2015} calculated the multipolar luminosity for vacuum environment and an extension of this, similar to \citet{Spitkovsky2006}, should be ideal for this study.

The angular dependence and scaling laws of multipolar torque in presence of magnetosphere remain an open numerical problem and may depend sensitively on the current distribution and field line topology. So we simply explore the case of multipolar field without magnetosphere and compare with a dipole field without magnetosphere (e.g., Figure \ref{fig:wdtorpol}).

The electromagnetic luminosity due to all multipolar fields $(\ell, m)$ is given by
\begin{equation}
L_\ell^{(m)} = \sum _{l,m}f(\ell, m)\, \frac{B^2_l \sin^2\chi \, \Omega^{2\ell + 2} R^{2\ell + 4}}{c^{2\ell + 1}},
\label{eq:approx_Llm}
\end{equation}
where the coefficient $f(\ell, m)$ captures the angular structure and relative strength of each $(\ell, m)$ component, and is well approximated from \citealt{Petri2015} by the empirical fit as
\begin{align}
f(\ell, m) &= -1.48704 
- 0.661679\, \ell 
+ 3.99237\, m 
- 0.317783\, \ell^2 \nonumber \\
&\quad + 1.66028\, \ell m 
- 2.99547\, m^2 
+ 0.00132135\, \ell^3 \nonumber \\
&\quad + 0.412289\, \ell^2 m 
- 1.721\, \ell m^2 
+ 1.78338\, m^3.
\end{align}

\begin{figure}
\begin{center}
\includegraphics[width=\columnwidth]{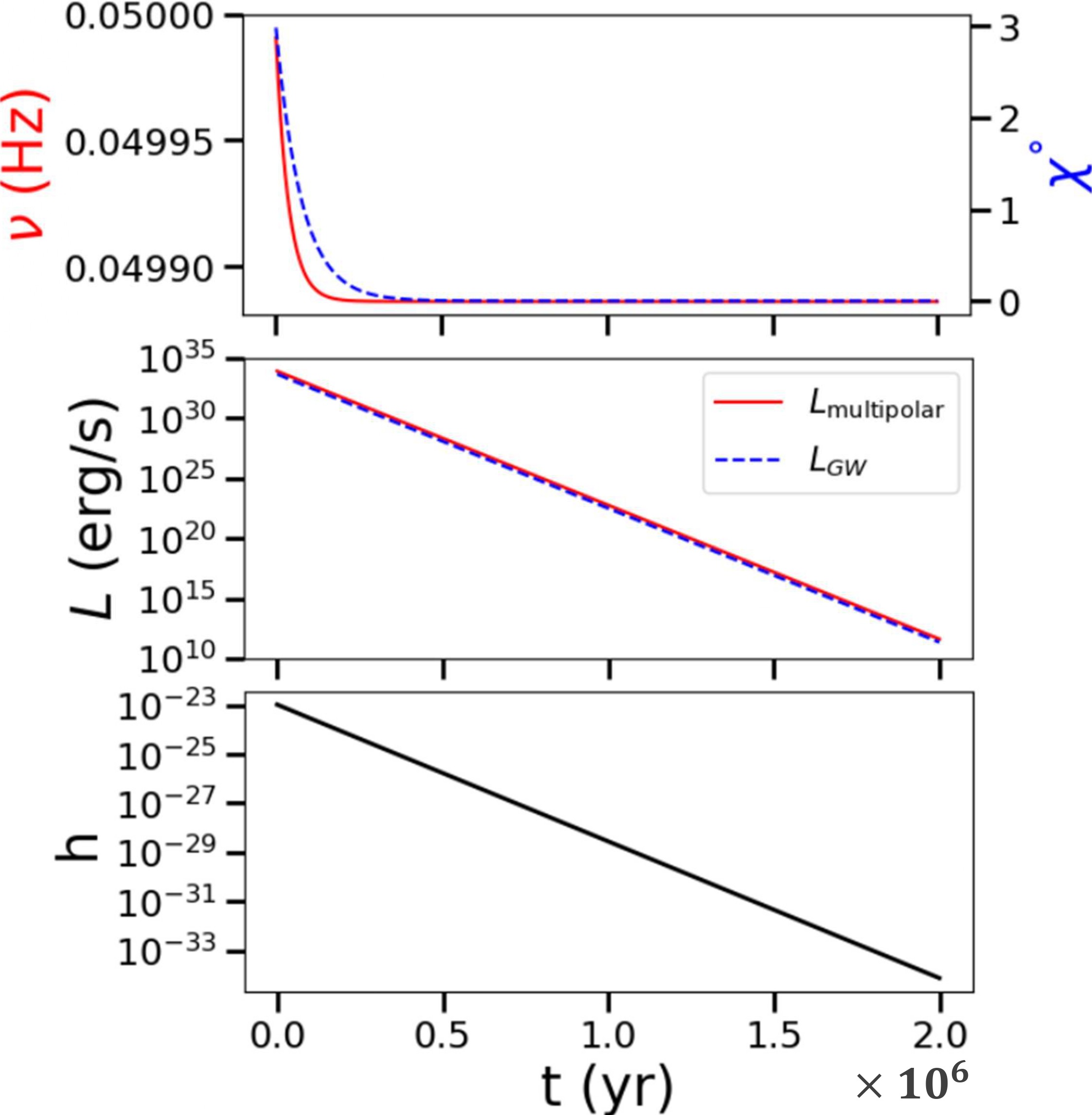}
\caption{Same as Figure \ref{fig:wdtorpol}, except if the WD1 has subdominant nonzero multipolar field.}
\label{fig:multpol}
\end{center}
\end{figure}

Replacing the dipolar radiation term in Equations~\eqref{eq:dwdt} and~\eqref{eq:dxdt}, the spin and obliquity evolution equations in the presence of magnetic multipoles with vacuum surrounding stands as
\begin{align}
\frac{d}{dt}(\Omega I_{z'z'}) 
&= -\frac{2G}{5c^5} (I_{zz} - I_{xx})^2 \, \Omega^5 \, \sin^2 \chi (1 + 15 \sin^2 \chi) \nonumber \\
&\quad - \sum _{l,m}f(\ell, m)\, \frac{B^2_l \Omega^{2\ell+1} R^{2\ell+4}}{c^{2\ell+1}} \sin^2 \chi,
\label{eq:general_dwdt}
\end{align}
and
\begin{align}
I_{z'z'} \frac{d\chi}{dt} 
&= -\frac{12G}{5c^5} (I_{zz} - I_{xx})^2 \, \Omega^4 \, \sin^3 \chi \cos \chi \nonumber \\
&\quad - \sum _{l,m} f(\ell, m)\, \frac{B^2_l \Omega^{2\ell} R^{2\ell+4}}{c^{2\ell+1}} \sin \chi \cos \chi,
\label{eq:general_dxdt}
\end{align}
where the effective surface field strength $B_l$ represents $l^{\rm th}$ order multipolar field at the stellar surface. Only the non-axisymmetric modes ($m \ne 0$) contribute to the radiated electromagnetic luminosity, as the axisymmetric components decay rapidly and do not produce outgoing radiation. The torque and spin-down luminosity retain angular dependencies similar to the dipole case \citep{CH1970}, but are modulated by mode-dependent factors $f(\ell, m)$ determined by the multipolar structure \citep{Petri2015}.

Notably, higher-order multipoles contribute less unless the magnetic field strength is at least two orders of magnitude greater than that of the preceding multipole. Similar to Section~\ref{dipolardecay}, we assume a quadrupolar magnetic field component at the surface of WD1, where $B_{\text{quad}} = 4 \times 10^{10}$\,G, which can be one to two order of magnitude larger than the dipolar component e.g. $B_p = 8 \times 10^8$\,G. Such a field, in internally toroidally dominated magnetic WDs, is neither dominant nor structurally impactful. This is consistent with observations suggesting that local multipolar magnetic fields—often an order of magnitude stronger than the average dipolar component—are required to explain Zeeman splitting in WDs \citep{Euchner2002,euchner2005}.

We study the angular momentum decay due to dipole and quadrupole ($m = 1, 2$) modes, as the dominance of higher mode multipoles would require unphysically strong magnetic fields. Figure~\ref{fig:multpol} shows the decays of $\nu$ and $\chi$; the middle panel shows the multipolar magnetic and gravitational luminosities for a WD1 like WD with multipolar field and no magnetosphere, obtained by solving Equations~\eqref{eq:general_dwdt} and \eqref{eq:general_dxdt} simultaneously. The corresponding GW amplitude decay is shown in the bottom panel of Figure~\ref{fig:multpol}. In this case, as the quadrupolar contribution dominates over the dipolar and GW radiations, the decay proceeds slightly faster (as $L_{\text{multipolar}} \gtrsim L_{{GW}}$) than the corresponding vacuum dipolar WD, see Figure \ref{fig:wdtorpol} where $L_{GW}\gtrsim L_D$. As this multipolar field provides similar decay timescale as high dipolar field (slightly higher $B_P$ than WD1+$B_P$ model of Figure \ref{fig:wdtorpol}), it will not be ventured extensively in next sections. Rather we will highlight the dipolar field with magnetosphere results for the most realistic WD configuration WD1+$B_P$, which is supposed to be stable magnetically and its consequences as it is the most realistic model in hand.

\section{GW detectability}
\label{GWdetect}

\begin{figure}
\begin{center}
\includegraphics[width=\columnwidth]{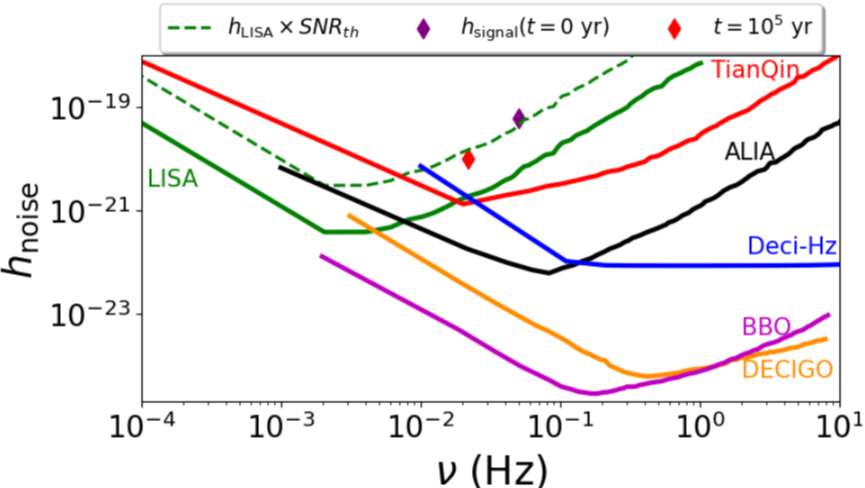}
\caption{Characteristic strain using Equation (\ref{eq:h_char}) for WDs like WD1+$B_P$ ($B_{max}^{Toroidal}=4\times10^{13}$ G, $B_{p}^{Poloidal}=8\times 10^{8}$ G, initial $\nu=0.05$ Hz and initial $\chi=3^\circ$, $d=1$ kpc) at young ($t\sim 0$) and old (near the end of active timescale, $t=10^5$ yr) ages are shown along with detector sensitivities.}
\label{fig:wdsens}
\end{center}
\end{figure}

To assess the detectability of these WDs -- WD1+$B_P$ with magnetosphere for instance, we compare their characteristic strain $h_{\text{signal}}$ with the sensitivity $h_{\text{noise}}$ of several planned space-based GW detectors—LISA, ALIA, DECIGO, Deci-Hz, BBO, and TianQin—as shown in Figure~\ref{fig:wdsens} (solid lines). For LISA, we also include $\text{SNR}_{\text{th}}\times h_{\text{signal}}$ (dashed line) to understand detection capability of LISA to detect that signal, where $\text{SNR}_{\text{th}} \sim 8$, is the typical signal-to-noise (SNR) threshold for LISA \citep{Tang2024}.
For the continuous search, $h_{\text{signal}}$ is calculated by \citet{Moore2015} as
\begin{equation}
h_{\text{signal}}=\sqrt{N_{\text{c}}}\left\langle h \right\rangle,
    \label{eq:h_char}
\end{equation}
where $N_{\text{c}}={\nu t}$, is the number of cycles from which the GW strain is collected and stacked by the detector over the whole observational time $T_{\text{obs}}=4$ yr,
and $\left\langle h \right\rangle$ is given by Equation \ref{eq:h_chi_avg}. Figure~\ref{fig:wdsens} shows that such highly magnetized young WDs should be detectable by BBO, DECIGO, Deci-Hz, TianQin, ALIA, and LISA 4-yr. However, $h_{\text{signal}}$ decays below LISA 4-yr detection capability, if the sources are older than $10^5$ yrs marking the end of its detectability window. We define this duration as the `active timescale' of CGW emission—beyond which such WDs are likely to become undetectable by LISA in 4 years.

\subsection{Signal to noise ratio}
\label{snr}

The SNR is a fundamental quantity that quantifies how well a GW signal can be distinguished from the detector’s background noise. A higher SNR implies a more confident detection. Therefore, calculating SNR is essential for assessing whether a signal from a WD is detectable by a given detector.

The challenges for detecting GWs from WDs using LISA are significant, particularly due to the weak strain amplitude of continuous signals. However, the probability of detection can be greatly enhanced by computing the SNR cumulatively over a long observational timescale.

In the simplest case, when the GW signal is constant over a short timescale, SNR is given by the ratio:
\begin{equation}
    \text{SNR} = \frac{h_{\text{signal}}}{h_{\text{noise}}},
\end{equation}
where $h_{\text{noise}}=\sqrt{\nu S_n(\nu)}$, $S_n$ is detector’s power spectral density (PSD).

While this expression holds for short-duration burst signals, it is not applicable to CGWs, which are long-lived and evolve over time as $\nu$ and $\chi$ gradually change. To accommodate this evolution, the total observation time $T_{\text{obs}}$ is divided into $\mathcal{N}$ number of shorter time stacks $T_{\text{s}}=T_{\text{obs}}/\mathcal{N}$, each sufficiently small so that $\nu$ and $\chi$ can be considered approximately constant within each segment. SNR is then built cumulatively by summing the squared contributions from each segment, as the signal phases do not align across the full observation window. This way the phase information between different stacks gets lost and thus it is called incoherent search.

For long-duration observations, a fully coherent search across the entire time span is computationally intensive, even if the signal amplitude remains constant. In contrast, an incoherent search using a stacked time-segment method offers a more practical and computationally efficient alternative \citep{brady2000, cutler2005}. Also, incoherent searches are essential since detectors can not maintain continuous good data due to interruptions and noise. If $\nu$ and $\chi$ change slowly or remain almost constant in each stack, the contribution from each stack can be calculated coherently and then added incoherently, this method can be called Semi-Coherent. This strategy is especially useful for all-sky searches targeting unknown pulsars, as demonstrated by \citet{MAG2008}. Also Doppler effects causing frequency modulation in continuous signals is inevitable. However, using small time stacks we can minimize this drift, allowing Doppler shifts to be neglected within each stack and avoiding complex demodulation. Adopting the semi-coherent stacking framework, the cumulative SNR is given by \citet{Bennett2010, KMTBM2021} as
\begin{equation}
\langle \text{SNR} \rangle = \sqrt{ \langle \text{SNR}_\nu^2 \rangle + \langle \text{SNR}_{2 \nu}^2 \rangle }.
\label{eq:snr}
\end{equation}
The total squared SNRs in Equation~\eqref{eq:snr} are obtained by summing the per-stack contributions at $\nu$ and $2\nu$ over all $\mathcal{N}$ stacks so that for a given $T_{\text{obs}}$ the sum runs over all stacks accumulated up to that time. They are given by
\citep{Jaranowski1998, Bennett2010}
\begin{equation}
\langle \text{SNR}_\nu^2 \rangle =  \frac{\sin^2 \zeta}{100} \, \sum_{s=1}^{\mathcal{N}}\frac{ T_{\text{s}} \, h_0^2\, \sin^2 2\chi}{S_n(\nu)},
\label{eq:snrw}
\end{equation}
\begin{equation}
\langle \text{SNR}_{2 \nu}^2 \rangle = \frac{4 \,\sin^2 \zeta}{25} \, \sum_{s=1}^{\mathcal{N}} \frac{T_{\text{s}} \, h_0^2 \, \sin^4 \chi}{S_n(2\nu)}.
\label{eq:snr2w}
\end{equation}
Here, $\zeta$ is the angle between the interferometer’s arms: $\zeta = 60^\circ$ for space-based detectors like LISA. The PSD data $S_n(\nu)$ are taken from publicly available sensitivity curves (e.g., \citealt{Moore2015}:
\url{https://gwplotter.com/}).

\begin{figure}
\begin{center}
\includegraphics[width=\columnwidth]{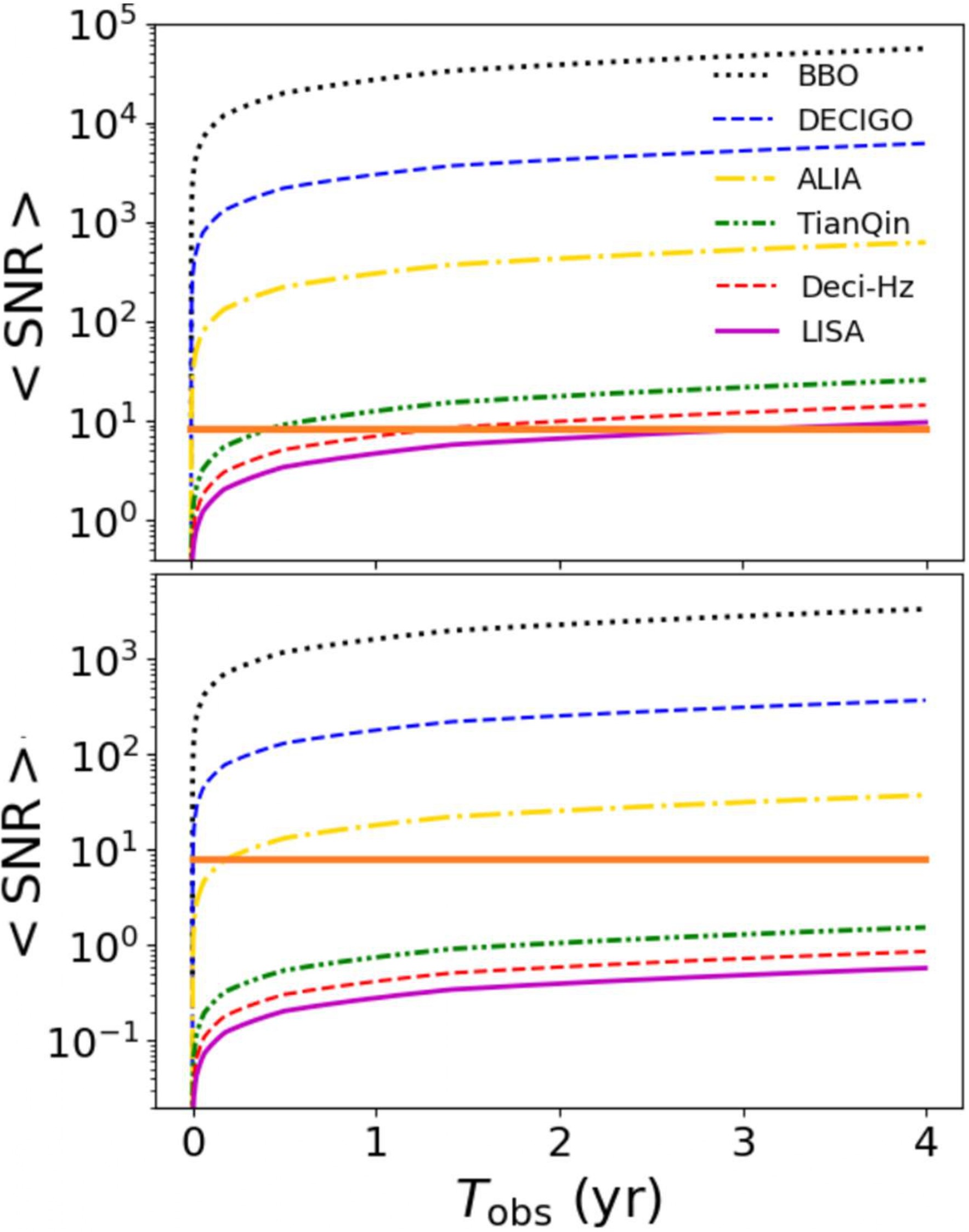}
\caption{Top Panel: SNR of WD1+$B_P$ (of $M=1.83\,M_\odot)$ for various detectors as a function of integration time assuming $d=1$kpc. Bottom Panel: Same as top panel except for WD3 (of $M=1.42\,M_\odot)$} -- a WD with weaker toroidal magnetic field. The orange line corresponds to $ \text{SNR}_{\text{th}}=8$.
\label{fig:snr}
\end{center}
\end{figure}

For a fully coherent search, $\text{SNR} \propto \sqrt{T_{\text{obs}}}$. In contrast, for a semi-coherent search the SNR in each stack scales as $\sqrt{T_{\text{s}}}$, since the signal is coherent within each stack and the contributions are then combined. If $h_0$, $\chi$, and $\nu$ remained constant, this would reduce to $\text{SNR} \propto \sqrt{\sum T_{\text{s}}} = \sqrt{T_{\text{obs}}}$, as expected.

In Figure~\ref{fig:snr}, we present the cumulative SNR up to continuous observation of 4 years for the toroidally dominated WD, WD1+$B_P$. Following \citet{Tang2024}, we adopt an SNR threshold of approximately 8 (i.e., $\geq 7$) to ensure a 95\% detection confidence after a 4 year mission. WD1+$B_P$ would be immediately detectable by BBO, DECIGO, and ALIA, whereas LISA, Deci-Hz and TianQin would require a few months of continuous integration to reach the detection threshold.

SNR initially grows as the GW amplitude accumulates over early observation intervals. However, as time progresses, $\nu$, $\chi$ and, consequently, $h$ diminish, leading SNR to eventually saturate. WDs' decay timescale much larger than $T_{\text{obs}}$ allows this accumulation to continue for longer, increasing the chance of surpassing the detection threshold at later stages. Slowly rotating WDs tend to decay more gradually, offering enhanced possibility of crossing the threshold. Meanwhile, highly magnetized WDs exhibit larger $h$ values, resulting in higher SNRs, which enhance their detectability, as shown in Figure~\ref{fig:snr}. In contrast, weakly magnetized WDs such as WD3 produce much lower SNRs (Figure~\ref{fig:snr}), often falling below the sensitivity thresholds of many detectors such as LISA, TianQin, Deci-Hz.

\subsection{Detection probability: Number of WDs}
\label{Detection plausibility}

In this section, we estimate the number of highly magnetized WDs that LISA will be able to detect. The classes of WDs, such as WD1+$B_P$, have an active timescale of $\sim 10^5$ yr according to our simulations, within which their GW strain remains above the detection threshold. Beyond this timescale, the GW amplitude becomes too weak for detection (see Figure \ref{fig:wdsens}). Therefore, only young, highly magnetized WDs are viable CGW sources.

The number of magnetized WDs to be detectable by LISA once active, denoted as $N_{\text{MWDs}}$, is given by
\begin{align}
    N_{\text{MWDs}} 
    &= p\% \times N_{\text{WD}} \notag \\
    &= p\% \times \text{birthrate yr\textsuperscript{-1}pc\textsuperscript{-3}} \times V~(\text{pc\textsuperscript{3}}) \notag \\
    &\quad \times \text{active timescale (yr)},
    \label{eq:nowd}
\end{align}
where $N_{\text{MWDs}}$ and $N_{\text{WDs}}$ represent the number of (highly) magnetized and total WDs, respectively. $V$ is the volume inside galaxy in which LISA is capable of searching for the magnetized WDs. The factor $p\%$ denotes the fraction of highly magnetized WDs relative to all WDs. The quantities $p$, birthrate, and active timescale all depend on specific assumptions about the formation and evolution channels of highly magnetized, rapidly rotating WDs. In particular, $p$ and the birthrate depend on whether these systems originate through the single-degenerate or double-degenerate channel. The active timescale is derived from XNS simulations and varies with the WD model: larger WDs have greater magnetic moments and moments of inertia, leading to faster decay of $\nu$ and $\chi$ (see Equations \ref{eq:dwdt}-\ref{eq:dxdt}). It also depends on the presence of a magnetosphere (see Section~\ref{spindown}) and on the assumed surface poloidal field strength. Since WD1+$B_P$ appears to be the most realistic configuration, we focus our discussion on WDs resembling that model.

\begin{figure}
\begin{center}
\includegraphics[width=\columnwidth]{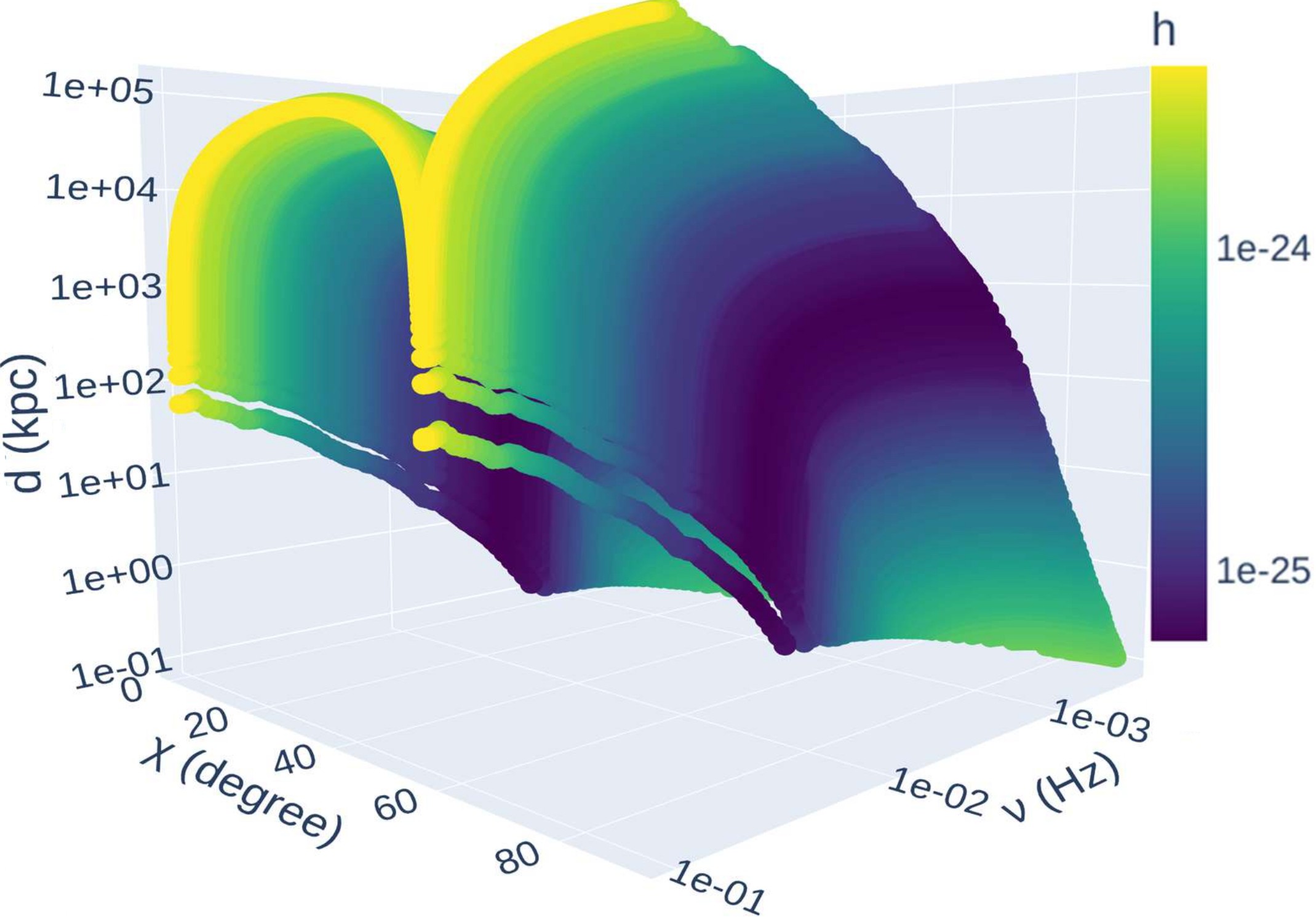}
\caption{
Maximum detectable distance $d$ by LISA for the model WD1+$B_P$ with $B_{\mathrm{max}}^{\mathrm{Toroidal}} = 4 \times 10^{13}\,$G and $B_{p}^{\mathrm{Poloidal}} = 8 \times 10^{8}\,$G, shown as a function of spin frequency $\nu$ and magnetic inclination angle $\chi$. The detector sensitivity $h$ is indicated by the color scale. Interactive versions of this figure are available at \href{https://drive.google.com/file/d/1xTNrfHjfoxolKnU59u7ANJs9G6d0_L79/view?usp=sharing}{this link (video)}. Also  \href{https://drive.google.com/file/d/1GQ9f50kV2C1m8kOwNSdMf5p_56WA48Nf/view?usp=sharing}{download HTML here.}}
\label{fig:3dlisa}
\end{center}
\end{figure}

\begin{figure}
\begin{center}
\includegraphics[width=\columnwidth]{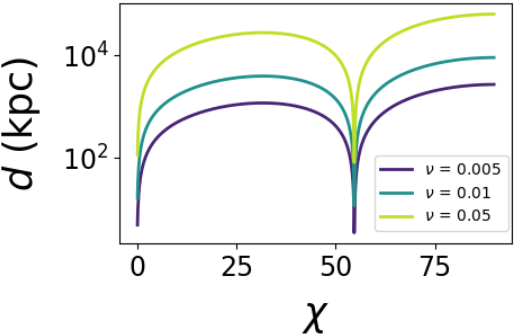}
\caption{A 2D representation of Figure \ref{fig:3dlisa}: variations of $d$ with $\chi$ for various $\nu$.}
\label{fig:3d2dlisa}
\end{center}
\end{figure}

To evaluate $N_{\text{MWDs}}$, we must determine the distance $d$ and thus volume $V$ up to which LISA can detect magnetized WDs, limited by its sensitivity. We consider a representative WD population with magnetic field strength similar to WD1+$B_P$ but rotating across a frequency range of $10^{-1}$--$10^{-5}$ Hz (as observed in \citealt{Kawka2020}) and spanning all possible $\chi$.

All such WDs should be detectable out to a maximum distance $d$ where their signal for $T_{\text{obs}}=4$ yr matches the LISA's detector sensitivity:
\begin{align}
    h_{\text{noise}}(\nu) &= \frac{h_{\text{signal}}}{\text{SNR}_{\text{th}}} = \left\langle h(\nu,\chi) \right\rangle_i \frac{\sqrt{N_{\text{c}}}}{\text{SNR}_{\text{th}}}.
    \label{eq:h_detector}
\end{align}
Using the inclination-averaged GW strain from Equation~\eqref{eq:h_chi_avg}, we can write
\begin{align}
    h_{\text{noise}}(\nu) &= f(\chi)\frac{2G}{c^4} \frac{4\pi^2\nu^2 \epsilon I_{xx}}{d} \frac{\sqrt{\nu t}}{\text{SNR}_{\text{th}}}.
    \label{eq:h_detector_fchi}
\end{align}
Thus, $d$ can be calculated for each $\nu$, $\chi$ when $h_{\text{noise}}$ and $\epsilon$ from the simulated WD model are known, i.e.,
\begin{align}
    d(\nu,\chi) &= f(\chi)\frac{2G}{c^4} \frac{4\pi^2\nu^2 \epsilon I_{xx}}{h_{\text{noise}}(\nu)}\frac{\sqrt{\nu t}}{\text{SNR}_{\text{th}}}.
    \label{eq:d_fchi}
\end{align}
Accordingly, we show in Figure~\ref{fig:3dlisa} the maximum detectable distance $d$ by LISA for WD1+$B_P$-like WDs with ranges of $\nu$ and $\chi$. Also Figure~\ref{fig:3d2dlisa} exhibits a 2D counterpart of it for three different values of $\nu$.

\begin{figure}
\begin{center}
\includegraphics[width=\columnwidth]{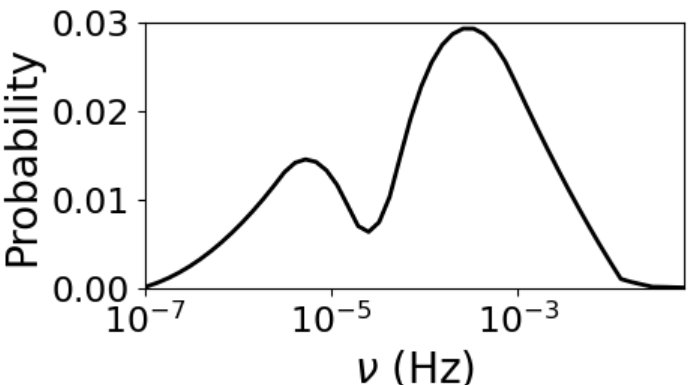}
\caption{Observed probability distribution of WD's various spin frequency from \citet{Kawka2020}.}
\label{fig:pdex}
\end{center}
\end{figure}

We now compute the average $d$ over all possible values of $\chi$ and $\nu$. First, we average over all possible $\chi$ as
\begin{align}
\left\langle d(\nu) \right\rangle_{\chi}
&= \frac{2G}{c^4} \frac{4\pi^2\nu^2 \epsilon I_{xx}}{h_{\text{detector}}(\nu)}\frac{\sqrt{\nu t}}{\text{SNR}_{\text{th}}}
\int_0^{\pi/2} f(\chi)\, \sin \chi \, d\chi,
\label{eq:d_avg_chi}
\end{align}
when the factor $\int_0^{\pi/2} f(\chi)\, \sin \chi \, d\chi=0.147$. Then we compute expected value of distance weighted by the observed probability distribution \(P(\nu)\) over $\nu$ using the corresponding $h_{\text{detector}}(\nu)$ sensitivity data of LISA. The function \(P(\nu)\) over WD's $\nu$ is inferred from the observed sample of magnetic WDs as reported by \citet{Kawka2020} based on the spin period statistics and shown in Figure \ref{fig:pdex}. We assume that the strong field does not affect the distribution. We thence obtain the averaged distance as
\begin{align}
\left\langle d \right\rangle_{\nu,\chi}
&= \sum P(\nu)\left\langle d(\nu) \right\rangle_{\chi} \sim 120 \text{ kpc}.
\label{eq:d_avg_nuchi}
\end{align}
However, the calculated average detection distance $\left\langle d \right\rangle_{\nu,\chi}$ significantly exceeds the size of Milky-Way, whose disk has a diameter of $\mathcal{D} \sim30$ kpc and a thickness $l \sim 300$ pc. Since there is no substantial population of WDs beyond our Galaxy and the nearest galaxy (Andromeda) lies at a much greater distance, it is reasonable to take the cylindrical volume of our galaxy as the effective limit for GW detectability from magnetized WDs, which is
\begin{align}
V = \pi \left(\frac{\mathcal{D}}{2}\right)^2 l = \pi\times (15\times 10^3)^2\times 300 \approx 2\times 10^{11}\, \text{pc}^3.
\label{eq:vol}
\end{align}

To calculate $N_{\text{MWDs}}$, we must first determine $p$\%, which depends on the type of progenitors from which these highly magnetized WDs form. We, therefore, consider the two most likely formation channels: the single-degenerate (accreting WD) and double-degenerate (WD merger) scenarios.

\subsubsection{`$p$' from single-degenerate channel}
\label{singledg}

The WD birthrate is observed to be $2\times10^{-12}$ pc\textsuperscript{-3} yr\textsuperscript{-1} \citep{Holberg2016}. Following the argument that overluminous Type Ia supernovae {(e.g., \citealt{howell2006})} may originate from highly magnetized super-Chandrasekhar WDs in the single-degenerate scenario {\citep{dasmuk13,Nomoto2018}}, we estimate $p$ accordingly. Although exceeding the Chandrasekhar mass is not essential for CGW detection, as we show in the following Section~\ref{ztf}, the presence of strong internal magnetic fields and rapid rotation are essential for detectable CGW. Such fields can, in some cases, support masses above the Chandrasekhar limit, as in the WD1 model. 
Out of all type Ia supernovae, at most 0.3\% are overluminous \citep{Meng2011}, which can be originated from super-Chandrasekhar WDs \citep{dasmuk13,Das2014}. 

However, it is unclear whether these events arise predominantly through the single–degenerate channel, since no more than $\sim 20\%$ of type Ia supernovae are thought to originate from single–degenerate systems \citep{gonz2012}. In particular, overluminous events are often associated with the double-degenerate channel in the literature. Nevertheless, not all WD mergers yield explosions luminous enough to explain overluminous Type Ia supernovae, since many lead instead to normal-luminosity detonations, off-center ignition and collapse, or the formation of long-lived massive WD remnants \citep{Pakmor2012,SaioNomoto1985,Shen2012,Dan2014, Maoz2014,Sato2015,Schwab2016}. Therefore, $p$ can be estimated to be around 0.06\% of the entire WD population. We emphasize that this estimate refers to WDs that may subsequently accrete and act as progenitor of overluminous type Ia supernovae. {In other words, all rapidly rotating, strongly magnetised WDs considered here are assumed to have undergone or to be undergoing accretion in a binary system.} Rapid rotation and strong magnetic fields in super-Chandrasekhar candidates can naturally reflect such an accretion history (e.g. flux freezing), even if the systems now appear to be isolated. Here, accretion spins up and amplifies magnetic fields due to flux freezing of WDs \citep{cumming2002,tout2008,Mukhopadhyay2017}. However, if we remove the restriction of super-Chandrasekhar, considering only highly magnetized WDs, $p$ would be much larger.

Therefore, we calculate the number of magnetized WDs which are accumulated in $10^6$ yr and to be detected by LISA from Equation \eqref{eq:nowd} as
\begin{align}
    N_{\text{MWDs}}  &=  \left[p\%\right] \times  \left[\text{birthrate}\right] \times  \left[V\right] \times  \left[\text{active timescale}\right]  \notag \\
    &=\left[\frac{0.06}{100}\right] \times \left[2\times 10^{-12} ~(\text{pc\textsuperscript{-3} yr\textsuperscript{-1}})\right]\notag \\ & \quad\times  \left[ 2\times 10^{11}(\text{pc\textsuperscript{3}})\right] 
     \times \left[10^5\text{(yr)}\right] \notag\\
    & = 24.
    \label{eq:nowdcal}
\end{align}
This means that LISA could detect a few dozens of highly magnetized WDs across the Galaxy during its mission as soon as it starts detection. The detection of CGW from these highly magnetized WDs may confirm the existence of super-Chandrasekhar WDs resulting from high magnetic field as possible progenitor of peculiar overluminous type Ia supernovae like SN 2003fg \citep{howell2006}.

\subsubsection{`$p$' from double-degenerate channel}

Another possible and independent scenario is the double-degenerate merger channel \citealt{Ferrario05,GarciaBerro2012,ZTF}, which can also produce highly magnetized, rapidly rotating, and potentially massive (near-Chandrasekhar) WDs. \citet{Brown2016} reported a Galactic double WD merger rate $\sim3\times10^{-3}\,\mathrm{yr^{-1}}$, but this rate applies primarily to binaries containing at least one extremely low mass WD (of $M\lesssim0.3\,M_\odot$), which originates from a $\sim1\,M_\odot$ main-sequence progenitor. In contrast, the progenitors of $\sim0.7+0.7\,M_\odot$ WD binaries are typically main-sequence stars with 
$M \gtrsim 3.5\,M_\odot$ \citep{Hollands2024}. According to the Galactic initial mass function \citep{Kroupa2001}, such $\gtrsim3.5\,M_\odot$ progenitors are about five times less common than $1\,M_\odot$ progenitors. Applying this correction reduces the effective merger rate to $\sim6\times10^{-4}\,\mathrm{yr^{-1}}$, which is the birthrate of highly magnetized WDs via double-degenerate channel (equivalent to $p\% \times$ birthrate of all WDs in Equation \ref{eq:nowd}).
 Over an active timescale of $10^{5}\,\mathrm{yr}$, like the estimate done in Section \ref{singledg}, the expected number of remnants is $\sim 60$ from double-degenerate channel.

It should be noted, however, that simulations of double-WD mergers do not robustly produce long-lived massive WD remnants always, even for systems starting with near-Chandrasekhar components \citep{Shen2012,Dan2014}. Outcomes include prompt thermonuclear type Ia supernova in some violent mergers \citep{Pakmor2012}, or quiescent off-center carbon ignition often leading to collapse rather than a stable WD remnant \citep{SaioNomoto1985,Schwab2016}. Broadly summarizing, only $\sim 10$\% of double-degenerate mergers are expected to result in a thermonuclear type Ia supernova with no remnant left behind \citep{Maoz2014,Sato2015}, while the vast majority ($\sim$90\%) merge {without a supernova explosion}. The merger-based number is therefore regarded as approximately $60\times 0.9=54$ as the production rate of strongly magnetized remnants, with the precise value depending on the uncertain remnant fraction.\\

The number mentioned above should be interpreted as the production of strongly magnetized remnants rather than confirmed massive, even super-Chandrasekhar, WDs. Combining two approaches, single- and double-degenerate, suggest a detectable population in the range of $\sim 24{-}54$. Given the uncertainty in the fraction of double-degenerate mergers that explode as Type Ia, the merger-based estimate should be regarded as uncertain but of the same order of magnitude as the single-degenerate estimate.

\section{Detection of ZTF J1901+1458}
\label{ztf}

\begin{table*}
\begin{center}
\caption{Observed and simulated properties of ZTF J1901+1458 assuming $\chi=30\degree$}
\small
\begin{tabular}{ccccc|cccccccc}
\hline\hline
\multicolumn{5}{c|}{\textbf{Observed}} & \multicolumn{8}{c}{\textbf{Simulated}} \\
\hline
$M$ & $R$ & $B_{p}^{\text{Pol}}$ & $T\!=\!1/\nu$ & $d$ & $\rho_c$ & $M$ & $R_E$ & $B_{\text{max}}^{\text{Tor}}$ & $\nu$ & ME/GE & $|\epsilon|$ & $h$ \\
$(M_{\odot})$ & (km) & (G) & (s) & (pc) & (g cm\textsuperscript{-3}) & $(M_\odot)$ & (km) & (G) & (Hz) & & & \\
\hline
$1.327$--$1.365$ & $\sim2140$ \!& $7$--$9\! \times\! 10^8$ & 416.24 & $41.40\! \pm\! 0.08$ & $3.7\! \times\! 10^9$ & 1.358 & 2185 & $7\! \times\! 10^{13}$ & 0.0024 & 0.06 & 0.14 & $3\! \times\! 10^{-24}$ \\
\hline
\end{tabular}
\label{tab:ztfdata}
\end{center}
\end{table*}

Here we will study the detection possibility of recently discovered ZTF J1901+1458—one of the most compact, near-Chandrasekhar mass, rapidly rotating and strongly magnetized WDs \citep{ZTF} as CGW emitter for LISA mission. This WD stands out for targeted search as a potentially detectable source as it is highly magnetized and fast rotating compared to many other observed WDs.

\begin{figure}
\begin{center}
\includegraphics[width=0.9\columnwidth]{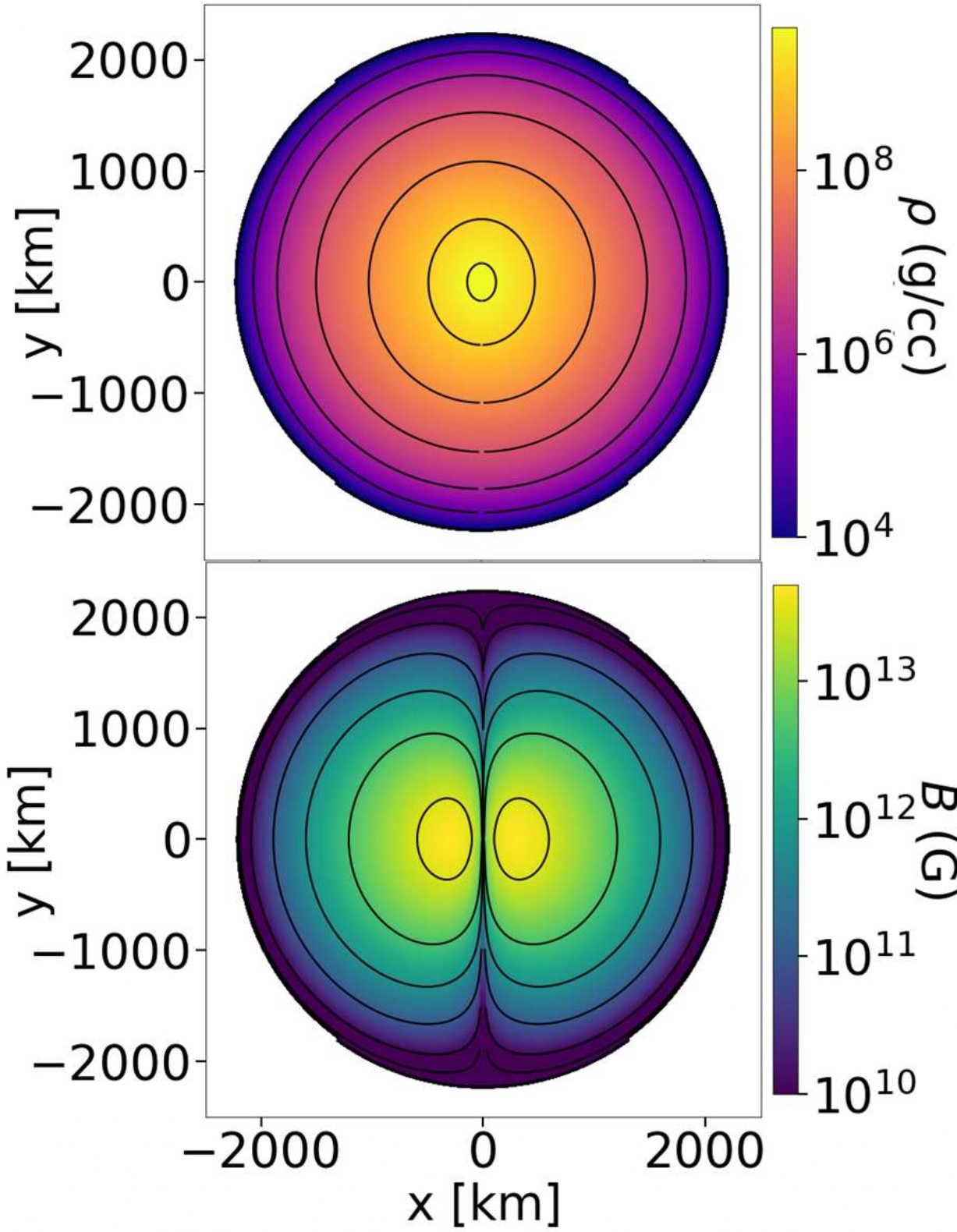}
\caption{Density isocontour (top panel) and magnetic field isocontour (bottom panel) of ZTF J1901+1458 model.}
\label{fig:xnsztf}
\end{center}
\end{figure}

To calculate the GW strain emitted by ZTF J1901+1458, we have to calculate the ellipticity and thus we model ZTF J1901+1458 with XNS. Following the previous sections, we construct a WD model based on the measured mass and rotation inferred by photometry, Gaia parallax from \citet{ZTF}. For the internal magnetic field, we choose values consistent with the surface field \citep{ZTF} inferred from Zeeman splitting (see Table~\ref{tab:ztfdata}). The rotation period is calculated from light curve variability, and the distance from Gaia parallax \citep{ZTF}. The simulated properties of the ZTF J1901+1458 model are also mentioned in the same table and the model is shown in Figure \ref{fig:xnsztf}. For a unique central density, the mass and radius are similar to those of the observed ones. The model has a rotation period of $416$ s and an internal toroidally dominated maximum magnetic field of $7\times 10^{13}$ G, which is chosen to be consistent with the observed dipolar field at the pole ($\sim 10^9$ G). As we have discussed that the stable stellar magnetic field configuration is supposed to consist of toroidal magnetic field, at least one order of magnitude higher than the poloidal counterpart. Hence, the toroidal field at surface should be $\sim 10^{10}$ G. We model accordingly to get such surface toroidal field and corresponding the maximum toroidal field comes out to be $\sim 7\times10^{13}$ G, see Figure \ref{fig:xnsztf} (bottom panel).

We compute $\epsilon$ from the WD model, enabling the computation of $\left\langle h(\nu,\chi) \right\rangle_i$ from Equation~\eqref{eq:h_chi_avg}, provided we know $\chi$ of ZTF J1901+1458. The observed photometric variation in this object is attributed to magnetic dichroism \citep{ZTF}, interpreted as rotational modulation rather than pulsation, where a strong magnetic field induces surface temperature and continuum opacity variations. As the WD rotates, these spatial inhomogeneities produce phase-dependent flux changes. A natural explanation for such dichroism is an oblique magnetic field, supporting the presence of a non-zero $\chi$ \citep{Putney1992,Suto2023}.

\begin{figure}
\begin{center}
\includegraphics[width=\columnwidth]{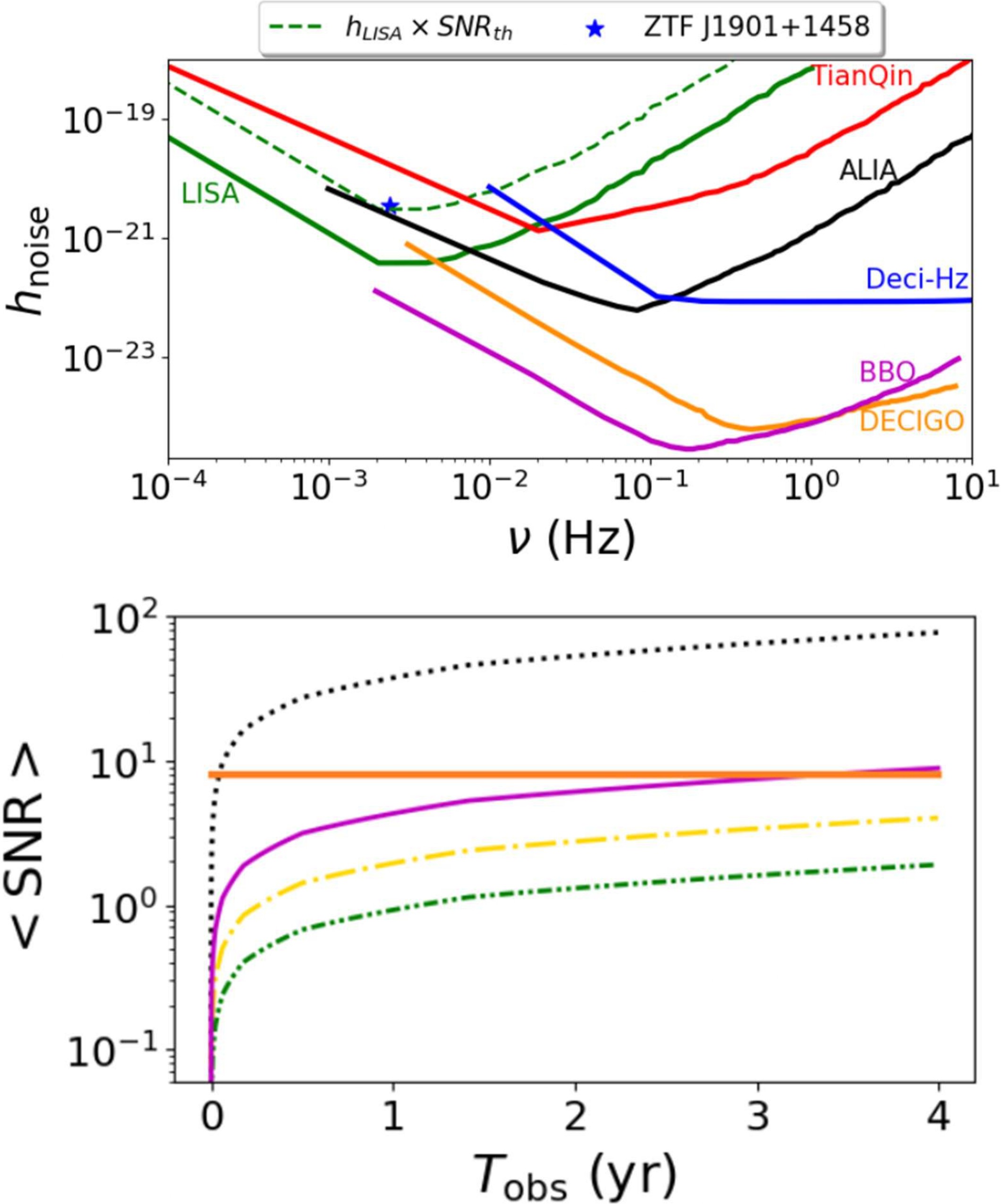}
\caption{Top panel: Characteristic strain from ZTF J1901+1458 for 4 year mission with present (initial) $\chi\sim 30\degree$ and $\nu=1/416$ Hz. Bottom panel: SNR of ZTF J1901+1458 for various detectors for 4 year long mission. The orange line corresponds to $\text{SNR}_{\text{th}}=8$. The SNRs are shown only for the detectors in the relevant frequency range.}
\label{fig:senztf}
\end{center}
\end{figure}

While the exact value of $\chi$ is not observationally constrained for ZTF J1901+1458, we proceed by assuming an optimistic $\chi=30\degree$. Thus, $h_{\text{signal}}$ of the object is shown alongside the sensitivity curves of various detectors in the top panel of Figure \ref{fig:senztf}. This suggests that ZTF J1901+1458 should be detectable by LISA over a 4 year observation, under the following assumptions: an optimistic choice of $\chi$, as well as specific internal magnetic field strength (or stronger than that) and its geometry. Even though this object rotates more slowly ($2.4\times10^{-3}$ Hz compared to $0.05$ Hz) and is less deformed (smaller $\epsilon$) than our previous model WD1, its GW strain $h$ remains detectable by LISA over 4 years owing to its much closer proximity (41 pc compared to 1 kpc). As we are not sure about exact $\chi$ of the object, the actual detectable signal may vary\footnote{Figure~\ref{fig:3d2dlisa} illustrates how the maximum detectable distance varies with $\chi$, similar to the variation of the signal with $\chi$, since both the signal and distance are proportional to $f(\chi)$.}. The cumulative SNR calculation also suggests that although ZTF J1901+1458—with initial $\chi \sim 30\degree$ and rotation frequency $\nu = 1/416$ Hz—will not be detected immediately, it becomes detectable after approximately 3.5 yr of continuous signal accumulation by LISA. This makes it one of the most promising candidates among highly magnetized WDs for detection by LISA, the most imminent next-generation detector. Such a detection would also provide critical insights into the internal magnetic field strength and geometry of the star.

\subsection{Distinguishing GWs from Other Sources}
\label{gwdist}

 In order to establish that a detected continuous GW signal originates from a rapidly rotating, magnetized WD and not from other compact binaries, it is necessary to distinguish our sources from alternative emitters in the same frequency band. In the $\sim$mHz range, persistent signals are also produced by AM\,CVn systems (ultracompact binaries with mass transfer) and detached double WD binaries, which dominate the Galactic population \citep{Lamberts2019}. These sources may therefore appear observationally similar to CGWs from isolated magnetized WDs. The key discriminant is the ability to resolve frequency derivatives ($\dot{\nu}$) in long observations \citep{Takahashi2002}, thereby separating isolated WDs from the binary foreground.

 AM\,CVns and DWDs exhibit orbital decay rates of $\lvert\dot \nu\rvert \sim 10^{-17}$--$10^{-16}$ Hz s\textsuperscript{-1} \citep{Gokhale2007} and up to $10^{-13}$ Hz s\textsuperscript{-1} \citep{Peters1964,Maggiore2008}, respectively, depending on orbital frequency.
Our WD1+$B_P$ model predicts $\lvert\dot \nu \rvert \sim 3\times 10^{-15}-10^{-14}$ Hz s\textsuperscript{-1} (see Figure~\ref{fig:wdtorpolmagsp}, top panel), depending on the age of WD. As we find the active timescale of the WD $10^5$ yr, $\lvert\dot \nu\rvert$ remains at least $3\times 10^{-15}$ Hz s\textsuperscript{-1} near the end of its active timescale. Thus $\lvert\dot \nu\rvert$ remains above LISA’s resolution $\Delta \dot \nu =4.3 /(\text{SNR}_{\text{th}}\,T_{\text{obs}}^2)\approx 3.4\times10^{-17}$ Hz s\textsuperscript{-1} for a 4 years observation with $\mathrm{SNR}=8$ \citep{Takahashi2002}. This would allow discrimination of our rapidly rotating WDs from AM\,CVns, though not necessarily from DWDs. However, for DWDs, the binary inspiral causes the orbital frequency to increase, resulting in a positive $\dot{\nu}$, in contrast to the negative $\dot{\nu}$ from our WD systems, which spin down continuously. The sign of $\dot{\nu}$ therefore provides a way to break the degeneracy between CGWs from DWDs and isolated WDs, provided both $\dot{\nu}$ values are measurable (i.e., $\lvert\dot{\nu}\rvert > \Delta \dot{\nu}$).
Nonetheless, for $\nu > 0.02$ Hz, the DWD population is essentially absent \citep{Lamberts2019}, so a CGW detected at such frequencies would almost certainly originate from a rapidly rotating magnetic WD.

For ZTF~J1901+1458, however, the expected $\lvert\dot{\nu}\rvert \sim 2\times10^{-20}\,\mathrm{Hz\,s^{-1}}$ lies below LISA’s resolution, appearing effectively monochromatic, so its GW signal alone cannot distinguish it from a DWD at the same frequency. In this case, electromagnetic constraints on rotation, magnetic field, and distance are essential for classification. 
This underscores the distinction between targeted sources like ZTF~J1901+1458, where electromagnetic priors break the degeneracy, and blind searches, where classification must rely solely on the resolvable $\dot \nu$. 
More generally, the ability to resolve $\dot \nu$ in matched-filter templates over a $4$ year baseline must be considered alongside frequency and SNR, since parameter-estimation uncertainties ultimately set the limits of source classification \citep{Cutler1994,Barack2004}.

\section{Conclusion}
\label{Conclusion}

We aim to investigate the plausibility of detecting magnetized WDs through CGWs using the upcoming space-based GW detector operating in the milliHertz band: LISA. By modeling the WDs using general relativistic magneto-hydrostatic configurations, we estimate the expected range of characteristic strain from such compact objects located at kiloparsec distances. However, the GW amplitude decays over time due to angular momentum loss via electromagnetic and quadrupole radiation. As a result, the GW emission weakens with age, and a realistic WD may no longer remain within the detectable sensitivity of LISA beyond a few million yr. This sets an upper detection timescale—an `active window'—implying that only relatively young, highly magnetized WDs are viable candidates for CGW detection.

We estimate that the number of such detectable WDs formed within this active timescale could be few dozens. A successful detection would not only constrain the internal magnetic field strength and configuration of these stars, but may also confirm the existence of highly magnetized WDs, one of the proposed progenitors of overluminous type Ia supernovae \citep{dasmuk13}, thereby offering an explanation for those peculiar supernovae.

As a caveat, our calculation for the single-degenerate channel treats these systems as effectively isolated (either companion might have been lost or accretion stops or negligible, at this moment) when estimating the active timescale. Systems with sustained strong accretion would require a different treatment of spin and magnetic-field evolution, which lies beyond the present scope. Such accreting WD systems {may} also emit GWs, but at orbital frequencies with very different $\nu$, $\dot\nu$, and amplitudes, making them distinguishable from the continuous signals considered here.

Additionally, we explore the case of ZTF J1901+1458—the most compact, near-Chandrasekhar mass, fast-rotating, and strongly magnetized WD discovered to date. Our simulations indicate that although this source may not be detectable immediately, it should be observed by LISA after approximately 3.5 yr of continuous integration during the targeted search, under the following assumptions: an optimistic choice of the obliquity angle, an internal magnetic field strength orders of magnitude stronger than the surface field, and a specific magnetic field geometry.

\section*{Acknowledgements}
MD expresses her heartfelt condolences on the passing of her masters supervisor {Somenath Chakraborty of Visva-Bharati University}, who introduced her to physics of Compact objects. The authors thank Rohan Raha (IISc) and Surajit Kalita (University of Warsaw) for fruitful discussions. The thanks are also due to the referee for the insightful and constructive comments, which have significantly improved the manuscript. MD is grateful to Soumallya Mitra, for his constant support and helpful conversations on probability calculations. She also acknowledges the Prime Minister’s Research Fellowship (PMRF) scheme, with Ref. No. TF/PMRF-22-5442.03.  BM acknowledges a project funded by SERB, India, with Ref. No. CRG/2022/003460, for partial support towards this research. TB acknowledges support  of the  National Science Center, Poland grant 2023/49/B/ST9/02777.

\bibliographystyle{aasjournal}
\bibliography{draft_m3.bib, reference} 

\end{document}